\documentclass[%
 reprint,
 amsmath,amssymb,
 aps,
]{revtex4-1}

\usepackage{float} 
\usepackage{hyperref}
\usepackage{graphicx,caption,subcaption}% Include figure files
\usepackage{dcolumn}% Align table columns on 
\usepackage{bm}% bold math
\usepackage{amssymb}
\usepackage{amsmath}
\usepackage{comment}
\usepackage{natbib}
\usepackage{physics}
\usepackage{tabu}
\usepackage{scrextend}
\usepackage{tabularx}
    \newcolumntype{L}{>{\raggedright\arraybackslash}X}

\newtheorem{theorem}{Theorem}[section]  

\begin{document}

\title{Asymmetric broadcasting of quantum correlations}% Force line breaks with \\

\author{Aditya Jain}
\affiliation{Center for Computational Natural Sciences and Bioinformatics, International Institute of Information Technology-Hyderabad, Gachibowli, Telangana-500032, India.} 
\author{Indranil Chakrabarty}
\affiliation{Center for Security, Theory and Algorithmic Research, International Institute of Information Technology-Hyderabad, Gachibowli, Telangana-500032, India.}
\author{Sourav Chatterjee}
\affiliation{Center for Computational Natural Sciences and Bioinformatics, International Institute of Information Technology-Hyderabad, Gachibowli, Telangana-500032, India.} 
\affiliation{SAOT, Erlangen Graduate School in Advanced Optical Technologies, Paul-Gordan-Strasse 6, 91052 Erlangen, Germany}

\date{\today}% It is always \today, today,
 
\begin{abstract}
In this work, we exhaustively investigate $1 \rightarrow 2$ local and nonlocal broadcasting of entanglement as well as correlations beyond entanglement (geometric discord) using asymmetric Pauli cloners with most general two qubit state as the resource. We exemplify asymmetric broadcasting of entanglement using \textsc{\char13}Maximally Entangled Mixed States\textsc{\char13}. We demonstrate the variation of broadcasting range with the amount of entanglement present in the resource state as well as with the asymmetry in the cloner. We show that it is impossible to optimally broadcast geometric discord with the help of these asymmetric Pauli cloning machines. We also study the problem of $1 \rightarrow 3$ broadcasting of entanglement using non-maximally entangled state (NME) as the resource. For this task, we introduce a method we call successive broadcasting which involves application of $1 \rightarrow 2$ optimal cloning machines multiple times. We compare and contrast the performance of this method with the application of direct $1 \rightarrow 3$ optimal cloning machines. We show that $1 \rightarrow 3$ optimal cloner does a better job at broadcasting than the successive application of $1 \rightarrow 2$ cloners and the successive method can be beneficial in the absence of $1 \rightarrow 3$ cloners. We also bring out the fundamental difference between the tasks of cloning and broadcasting in the final part of the manuscript. We create examples to show that there exist local unitaries which can be employed to give a better range for broadcasting. Such unitary operations are not only economical, but also surpass the best possible range obtained using existing cloning machines enabling broadcasting of lesser entangled states. This result opens up a new direction in exploration of methods to facilitate broadcasting which may outperform the standard strategies implemented through cloning transformations. 
\end{abstract}

\maketitle

%\tableofcontents

\section{Introduction}

In the last few decades, we have seen many instances where information processing with quantum resources either have outperformed their classical counterparts or may not have classical analogues. These tasks have created pioneering protocols with high impact in the domain of both communication and cryptography. In these domains, some of the popular protocols include teleportation \cite{ben1}, superdense coding \cite{ben21,ben22}, secret sharing \cite{secretsharing1,secretsharing2,secretsharing3,secretsharing4}, key distribution \cite{bb841}, digital signatures \cite{gottesmann,peunti} and remote state preparation \cite{remoteStateprep}. Few of them are even now commercially available \cite{idquantique}.

However, in the quantum world, the Heisenberg's uncertainty principle which arises from non-commutativity of mutually complementary operators, on one hand imposes strict limitations \cite{wootters, barnum-gen, janos}; whereas on the other hand becomes of paramount advantage \cite{bb841}. This leads to one of the famous \textsc{\char13}No-go\textsc{\char13} theorems \cite{wootters, buzek2,barnum,indranil2,lemm} in quantum mechanics called the No-cloning theorem \cite{wootters}. It states that simultaneous \textit{deterministic} and \textit{noiseless} cloning of an \textit{unknown} \textit{arbitrary} quantum state is impossible \cite{bruss}. In this context, \textit{deterministic} means that the cloning can be performed successfully on every input state, and, \textit{noiseless} implies that a pure state is transformed into another pure state and not a mixed state. Finally \textit{arbitrary} and \textit{unknown} impose the restrictions that the state cannot be chosen from any known probability distribution of input states (for instance, werner-state \cite{wernerstate} or bell-diagonal states \cite{lang}) and that the cloner's configuration must not be correlated with it \cite{bruss}. It is also equivalent to the statement that simultaneous cloning can happen only when two mixed states $\rho_1$, $\rho_2$ are orthogonal or identical \cite{barnum,lemm}.

First attempts towards the design of best approximate cloning methodology led to construction of Universal Quantum Cloning Machine (UQCM) \cite{buzek} possessing an input state independent fidelity of $\frac{5}{6}$. However, other instances where the cloning fidelity was a function of input state parameters \cite{bruss,adhikari2,muellerclone} or where a unit fidelity with non-zero success probability could be achieved, were also later developed \cite{duan1,hardy}. All such cloning machines were \textsc{\char13}symmetric\textsc{\char13} implying that the fidelity of the clones remained same and also \textsc{\char13}optimal\textsc{\char13} meaning that the fidelity achieved was maximal within the purview of quantum mechanics \cite{gisinmassar, bruss}. Later in \cite{cerf,ghiu,iblisdir} optimal asymmetric universal cloners that produce cloned outputs having different fidelities were also proposed. In this case, asymmetry refers to unequal distribution of information onto final systems from an initial system via a quantum evolutionary process and optimality represents the achievement of maximal fidelity on one output when the other is fixed \cite{zhang}. Asymmetric quantum cloning in general is useful particularly when analyzing the security of some quantum key distribution protocols. In such protocols the optimal eavesdropping strategy, when known, often turns out to coincide with optimal asymmetric cloning \cite{bruss,cerf} but this is not always the case \cite{ferenczi}. Other applications include telebroadcasting \cite{ghiu} and asymmetric telecloning \cite{murao,ghiu} of states which is useful when information needs to be distributed among parties with varying trust. 

Quantum entanglement lies at heart of quantum information and plays a pivotal role in all information processing protocols \cite{ben1,ben21,ben22,ekert,bb841,bb842,secretsharing1,secretsharing2,secretsharing3,secretsharing4}. Thus the aim to quantify the distribution of entanglement across various nodes (parties) in a network is of practical importance. A convenient way to achieve this task, primarily coined as \textsc{\char13}broadcasting of entanglement\textsc{\char13}, is through cloning \cite{buzek2}. In literature, for this purpose, state-dependent and independent approaches, with optimal or sub-optimal fidelities, in form of local or nonlocal (global) cloning transformations have been used \cite{adhikari,kar,manish,ghiu,sourav}. Broadcasting into more than two pairs in local case and more than six pairs in the nonlocal case is impossible with symmetric cloners \cite{kar}. We hypothesize that it might be possible to surpass the limit of six pairs using asymmetric cloners. First studies using optimal asymmetric cloners (local and nonlocal) to broadcast entanglement into two pairs was investigated in \cite{ghiu} with a non-maximally entangled input state. It was also shown that maximal broadcasting of entanglement is achieved only when asymmetric cloner reduces to a symmetric one \cite{ghiu}. However, there the limits for broadcasting of entanglement with larger class of mixed input states and considering maximization on the quantity of generated output pairs were not studied in details. Correlations also exist beyond entanglement, such as discord \cite{discord1, discord2} and have been discovered to be a valuable quantum resource in cryptography \cite{discordresource}. Broadcasting of quantum correlations beyond entanglement (QCsbE) has been explored with symmetric cloning operations \cite{sourav} but it still remains unexplored with asymmetric cloners. Such a question also gains practical importance in our context when on one hand we consider that asymmetric cloning directly relates out to major eavesdropping strategies \cite{ferenczi} and on the other hand take up discord as a useful resource in encrypted communication \cite{discordresource}. Apart from correlations, there have been attempts to broadcast other resources such as coherence \cite{udit} and more generally quantum states which had led to the famous no-broadcasting theorem \cite{barnum}. However, it is important to mention that here we restrict to the picture of approximate broadcasting of correlations via cloning and so we neither violate the no-broadcasting theorem \cite{barnum} nor the monogamy of entanglement \cite{ckw}.

In this work, firstly in Sec.~\ref{sec:def}, we define the required tools like Pauli asymmetric cloning machines, resources like the most general two-qubit mixed state \cite{gisinmix}, entanglement detection criteria and geometric discord. In Sec.~\ref{sec:1to2broadent}, after discussing the general result for two qubit mixed state, we exemplify the variation of the broadcasting range with the asymmetry of the cloner using Maximally Entangled Mixed States(MEMS). In Sec.~\ref{sec:1to2broaddis}, considering QCsbE (discord), we show that it is impossible to broadcast such correlations with these cloners optimally. In Secs.~\ref{sec:1to3broadentsucc} and~\ref{sec:1to3broadentdir}, we accomplish the task of $1 \rightarrow 3$ broadcasting of entanglement by employing two different strategies: the first strategy is a novel approach where we make successive use of $1 \rightarrow 2$ asymmetric cloners. This method has similarities with existing methods to implement quantum cloning by cellular automata \cite{automata}. They both involve successive application of unitaries. However, our approach is aimed at accomplishing a different task and is more scalable. The second strategy involves using $1 \rightarrow 3$ asymmetric cloner directly. We find that the second strategy i.e. the use of $1 \rightarrow 3$ asymmetric cloner performs better than the first one as far as the broadcasting range is concerned. However, for a large part of the input range, the successive approach might be preferred over the use of $1 \rightarrow 3$ cloners, which might be difficult to build in practice. Finally in Sec.~\ref{sec:arbunit}, we introduce for the first time the notion of broadcasting with arbitrary unitaries that extend beyond the domain of cloning machines. We demonstrate via numerical examples that broadcasting using arbitrary local unitaries instead of known cloning machines gives us a better range for broadcasting of entanglement. Another important advantage of this method is that it is economical i.e. does not employ ancillary qubits. This highlights a fundamental difference between the two tasks - cloning of states and broadcasting of entanglement. A summary of the previous results in contrast to the present work is given in Table \ref{Table1}.

\begin{table}[htbp]
{\bf $1 \rightarrow 2$ broadcasting of entanglement}
\begin{tabularx}{\linewidth}{ |c|c|L| }
\hline
{\bf Operation} & {\bf Resource state} & {\bf Author(s)}  \\ \hline
 Symmetric cloner & NME & Bu$\check z$ek {\it et al.} and Hillery \cite{buzek2,buzhil}    \\ \hline
Symmetric cloner& 2-qubit general & Chatterjee {\it et al.} \cite{sourav} \\ \hline
Asymmetric cloner& NME & Ghiu \cite{ghiu} \\ \hline
Asymmetric cloner& 2-qubit general & {\it present work [Sec.~\ref{sec:1to2broadent}]}  \\ \hline
Arbitrary unitary& Werner-like \& BDS & {\it present work [Sec.~\ref{sec:arbunit}]}  \\ \hline 
\end{tabularx}
\newline
\vspace*{0.1 cm}
\newline
{\bf $1 \rightarrow 2$ broadcasting of quantum discord}
\begin{tabularx}{\linewidth}{ |c|c|L| }
\hline
{\bf Cloning operation} & {\bf Resource state} & {\bf Author(s)}  \\ \hline
 Symmetric & 2-qubit general & Chatterjee {\it et al.} \cite{sourav}\\ \hline
 Asymmetric & 2-qubit general & {\it present work [Sec.~\ref{sec:1to2broaddis}]}    \\ \hline
\end{tabularx}
\newline
\vspace*{0.1 cm}
\newline
{\bf $1 \rightarrow 3$ broadcasting of entanglement}
\begin{tabularx}{\linewidth}{ |c|c|L| }
\hline
{\bf Cloning operation} & {\bf Resource state} & {\bf Author(s)}  \\ \hline
 $1\rightarrow3$ Symmetric & NME & Bandyopadhyay {\it et al.} \cite{kar}\\ \hline
 $1\rightarrow3$ Asymmetric & NME & {\it present work [Sec.~\ref{sec:1to3broadentdir}]}    \\ \hline
 $1\rightarrow2$ Asymmetric & NME & {\it present work [Sec.~\ref{sec:1to3broadentsucc}]}    \\ \hline
\end{tabularx}
\caption[]{Summary of earlier results and the present work on broadcasting of entanglement and discord. NME, BDS and 2-qubit general stand for Non maximally entangled state, Bell-diagonal states and general two-qubit mixed state respectively.}
\label{Table1}
\end{table}
\section{Useful definitions and related concepts}
\label{sec:def}
In this section, we introduce various concepts that are related to the central theme of the manuscript. 
\subsection{General two qubit mixed state}
In this artcile, we use a general two qubit mixed state (shared by parties numbered $1$ and $2$) as a resource state for some operations, and it is represented in the canonical form as \cite{gisinmix,sourav}
\begin{eqnarray}
\rho_{12}&=&\frac{1}{4}[\mathbb{I}_4+\sum_{i=1}^{3}(x_{i}\sigma_{i}\otimes \mathbb{I}_2+ y_{i}\mathbb{I}_2\otimes\sigma_{i})\nonumber\\
&+&\sum_{i,j=1}^{3}t_{ij}\sigma_{i}\otimes\sigma_{j}]=\{\vec{x},\:\vec{y},\: \mathbb{T}\}\:\:\: \mbox{(say),}
\label{eq:mix}
\end{eqnarray}

where $x_i=Tr[\rho_{12}(\sigma_{i}\otimes \mathbb{I}_2)]$, $y_i=Tr[\rho_{12}(\mathbb{I}_2\otimes\sigma_{i})]$ are local Bloch vectors. The correlation matrix is given by $\mathbb{T}=[t_{ij}]$ where $t_{ij}=Tr[\rho_{12}(\sigma_i\otimes\sigma_{j})]$ with [$\sigma_i;\:i$ = $\{1,2,3\}]$ being $2\otimes 2$ Pauli matrices and $\mathbb{I}_n$ is the identity matrix of order $n$.

\subsection{Detection of entanglement}
\label{sec:ph}
The Peres-Horodecki (\textbf{PH}) \cite{peres1,horodecki} criterion is used to test the separability of two quantum mechanical systems. This criteria is necessary and sufficient only for $2 \otimes 2$ and $2 \otimes 3$ dimensional bipartite systems. If at least one of the eigenvalues of a partially transposed density operator for a bipartite state $\rho_{12}$ defined as $\rho_{m\mu,n\nu}^{T}=\rho_{m\nu,n\mu}$ turn out to be negative, then we say that the state $\rho_{12}$ is entangled. Equivalently, this criterion can be translated to the condition that determinant of at least one of the two matrices
\begin{equation}
\begin{split}
\renewcommand{\arraystretch}{1.5} % give some more room
& W_{3}=
\left(\begin{array}{@{}c|c@{}}
  W_{2} &
  \begin{matrix}
  \rho_{00,10} \\
  \rho_{00,11} 
  \end{matrix}
\\ \hline
  \begin{matrix}
  \rho_{10,00} & \rho_{11,00}
  \end{matrix}
  & \rho_{10,10}
\end{array}\right)
 \:\:\text{and}\\
\renewcommand{\arraystretch}{1.5} % give some more room
& W_{4}=
\left(\begin{array}{@{}c|c@{}}
  W_{3} &
  \begin{matrix}
  \rho_{01,10} \\
  \rho_{01,11} \\
  \rho_{11,10} 
  \end{matrix}
\\ \hline
  \begin{matrix}
  \rho_{10,01} & \rho_{11,01} & \rho_{10,11} 
  \end{matrix}
  & \rho_{11,11} 
\end{array}\right)\:\:\:\:\:\:\:\:\:\:\:\:\:\:\:\:\:\:\:\:\:\:\:\:\:\:
\label{eq:w3w4}
\end{split}
\end{equation}

is negative; with determinant of 
%\begin{equation}
$W_2=\begin{bmatrix}\rho_{00,00} & \rho_{01,00} \\
\rho_{00,01} & \rho_{01,01} \\
\end{bmatrix}$ being simultaneously non-negative.
\subsection{Geometric Discord}
In the last decade, researchers came up with several measures to quantify the correlations in the quantum systems, that go beyond the standard notion of entanglement. In this article we have chosen geometric discord (GD) to quantify the quantum correlations beyond entanglement (QCsbE) \cite{discord1,discord2,discord3,discord4}. Geometric discord is a distance based measure defined for any general two qubit state $\rho_{12}$ (shared by parties numbered $1$ and $2$) as 
\begin{equation}
    \label{eq:gdiscord}
    D_{G}(\rho_{12}) = \min\limits_{\mathcal{X}}  ||\rho_{12} - \mathcal{X}||^{2},
\end{equation}
where $\mathcal{X}$ is the set of all classical states of the form $ p |\psi_1\rangle \langle \psi_1| \otimes \rho_1 + (1-p)|\psi_2\rangle \langle \psi_2| \otimes \rho_2 $. Here, $|\psi_1\rangle$ and $|\psi_2\rangle$ are two orthonormal basis of subsystem $1$. The states $\rho_1$ and $\rho_2$ are two density matrices of the subsystem $2$. $||\rho_{12} - \mathcal{X}||^2=\Tr ((\rho_{12} - \mathcal{X})^2)$ is the Hilbert-Schmidt quadratic norm. For an arbitrary two-qubit state, the expression for GD is as follows :
\begin{equation}
\label{eq:gd_arbitrary}
D_{G}(\rho_{12}) = \frac{1}{4}(||\vec{x}||^{2} + ||\mathbb{T}||^{2} - \lambda_{max}),
\end{equation}
where $||\mathbb{T}||^{2}=\Tr(\mathbb{T}^t \mathbb{T})$ and $\lambda_{max}$ is the maximal eigenvalue of the matrix $\Omega = (\vec{x}\vec{x}^{t} + \mathbb{T}\mathbb{T}^{t})$, where the superscript $t$ denotes transpose and $\mathbb{T}$ is the correlation matrix of $\rho_{12}$. 
\subsection{Asymmetric quantum cloning}
\subsubsection*{$1 \rightarrow 2$ cloning}
A $1\rightarrow 2$ cloning machine is said to be asymmetric when it takes $\rho$ and a blank state as inputs, and transforms them into two different copies $\rho_1$ and $\rho_2$. In this article, we use two kinds of optimal cloning machines. 
For broadcasting using local cloner, we use the optimal universal asymmetric Pauli cloning machine $\left(U^l_a\right)$ \cite{ghiu}
\begin{equation}
\begin{split}
& U^l_a|0\rangle|00\rangle=\frac{1}{\sqrt{1+p^2+q^2}}(|000\rangle+p|011\rangle+q|101\rangle),\\
& U^l_a|1\rangle|00\rangle=\frac{1}{\sqrt{1+p^2+q^2}}(|111\rangle+p|100\rangle+q|010\rangle),
\end{split}
\label{eq:local_cloner}
\end{equation}
where $p, q$ are the asymmetric parameters of the cloner and $p+q=1$. Having $p=q=1/2$ corresponds to a perfectly symmetric cloner. 

For nonlocal broadcasting, we use the nonlocal optimal universal asymmetric Pauli cloning machine $\left(U^{nl}_a\right)$ \cite{ghiu}
\begin{equation}
\begin{split}
& U^{nl}_a|j\rangle|00\rangle =\frac{1}{\sqrt{1+3(p^2+q^2)}}( |j\rangle|j\rangle|j\rangle \\
& + p\sum_{r=1}^{3}|j\rangle|\overline{j+r}\rangle|\overline{j+r}\rangle
+q\sum_{r=1}^{3}|\overline{j+r}\rangle|j\rangle|\overline{j+r}\rangle),
\end{split}
\label{eq:nonlocal_cloner}
\end{equation}
where $j \in \{0,1\}$ and $\overline{j+r}=(j+r)$ modulo $4$ and $p+q=1$. For both the cloners, the first two qubits represent the clones and the last qubit represents the ancillary qubit corresponding to the cloning machine. 
\subsubsection*{$1 \rightarrow 3$ cloning}
A $1\rightarrow 3$ cloning machine is said to be asymmetric when it takes $\rho$ and two blank states as inputs, and transforms them into three different copies $\rho_1$, $\rho_2$ and $\rho_3$. The optimal asymmetric cloning transformation that we use in this article is given by \cite{jaromir}
\begin{equation}
\begin{split}
&\ket{\psi} \rightarrow \mathcal{N} [\alpha \ket{\psi}_A (\ket{\Phi^+})_{BE}\ket{\Phi^+}_{CF}+\ket{\Phi^+})_{BF}\ket{\Phi^+})_{CE})\\
&+\beta \ket{\psi}_B (\ket{\Phi^+})_{AE}\ket{\Phi^+}_{CF}+\ket{\Phi^+})_{AF}\ket{\Phi^+})_{CE})\\
&+\gamma \ket{\psi}_C (\ket{\Phi^+})_{AE}\ket{\Phi^+}_{BF}+\ket{\Phi^+})_{AF}\ket{\Phi^+})_{BE})],
\end{split}
\label{eq:1to3cloner}
\end{equation}
where, labels A, B and C represent the states of the three clones, E and F represent ancilla;
$\mathcal{N}=\sqrt{\frac{d}{(2(d+1))}}$ is a normalization constant; 
$\ket{\Phi^+}=\frac{1}{\sqrt{d}}\sum\limits_{j=0}^{d-1}\ket{j}\ket{j}$ is the maximally entangled state; $j \in \{0,1\}$ and $d=dim(\mathcal{H}_{in})$ is the dimension of the input Hilbert space ($\mathcal{H}_{in}$). The asymmetric parameters $\alpha,\beta,\gamma$ satisfy the normalisation condition, $
\alpha^2 + \beta^2 + \gamma^2 + \frac{2}{d}(\alpha\beta+\beta\gamma+\alpha\gamma) = 1$.

We use the cloner with $d=2$ for local broadcasting and $d=4$ for nonlocal broadcasting. The values of the parameters for the perfectly symmetric case are $\alpha=\beta=\gamma=\frac{1}{\sqrt{6}}$ when $d=2$, and $\alpha=\beta=\gamma=\frac{\sqrt{2}}{3}$ when $d=4$.
\begin{figure}[htbp]
\centering
        \begin{subfigure}[b]{0.22\textwidth} \includegraphics[scale=0.25]{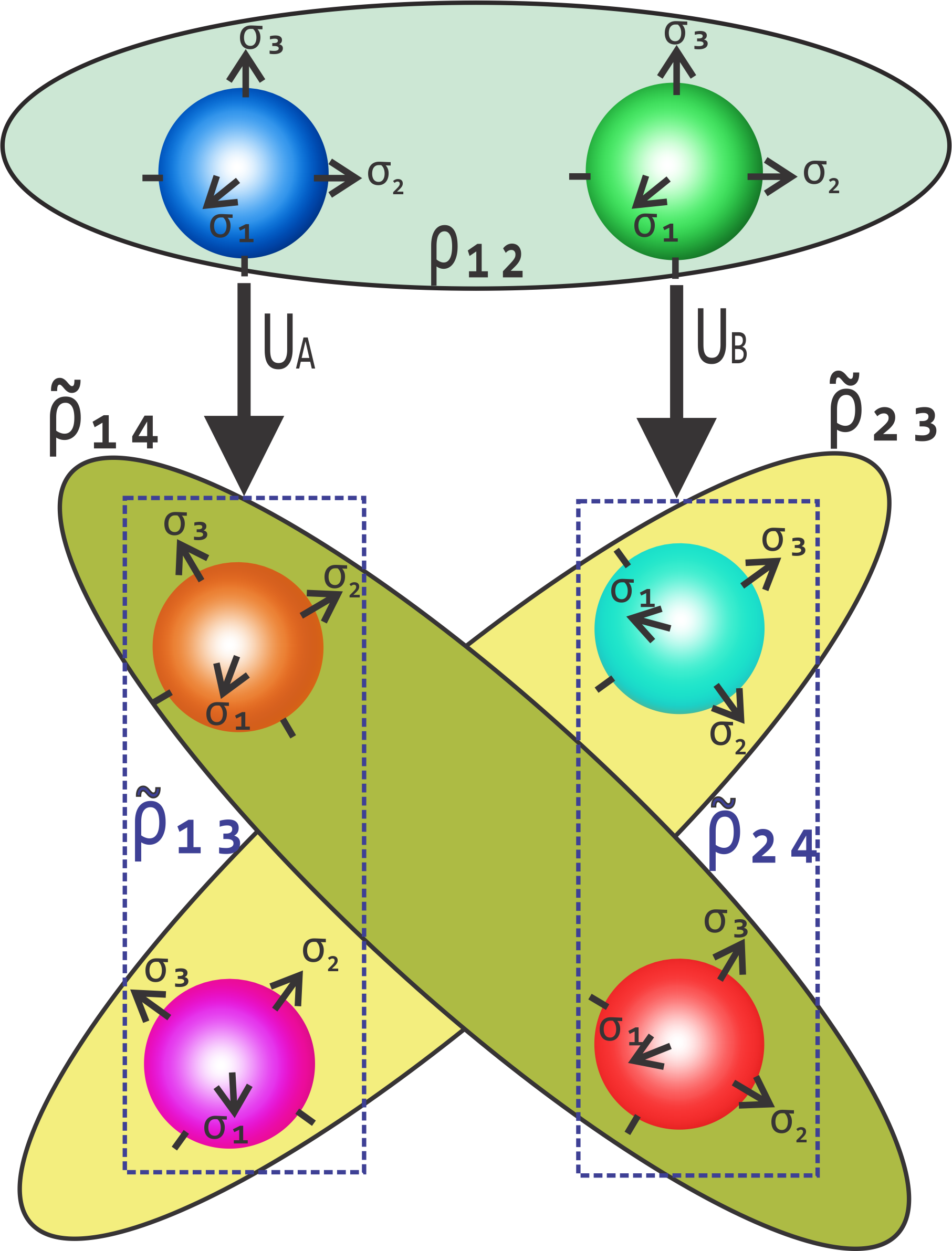}
                \caption{Local broadcasting. Only the diagonal nonlocal output pairs ($\tilde{\rho}_{14}$ and $\tilde{\rho}_{23}$) have been highlighted for clarity.}
                \label{fig:1to2loc}
        \end{subfigure}\hspace{0.02\textwidth}%
        \begin{subfigure}[b]{0.23\textwidth}
            \includegraphics[scale=0.25]{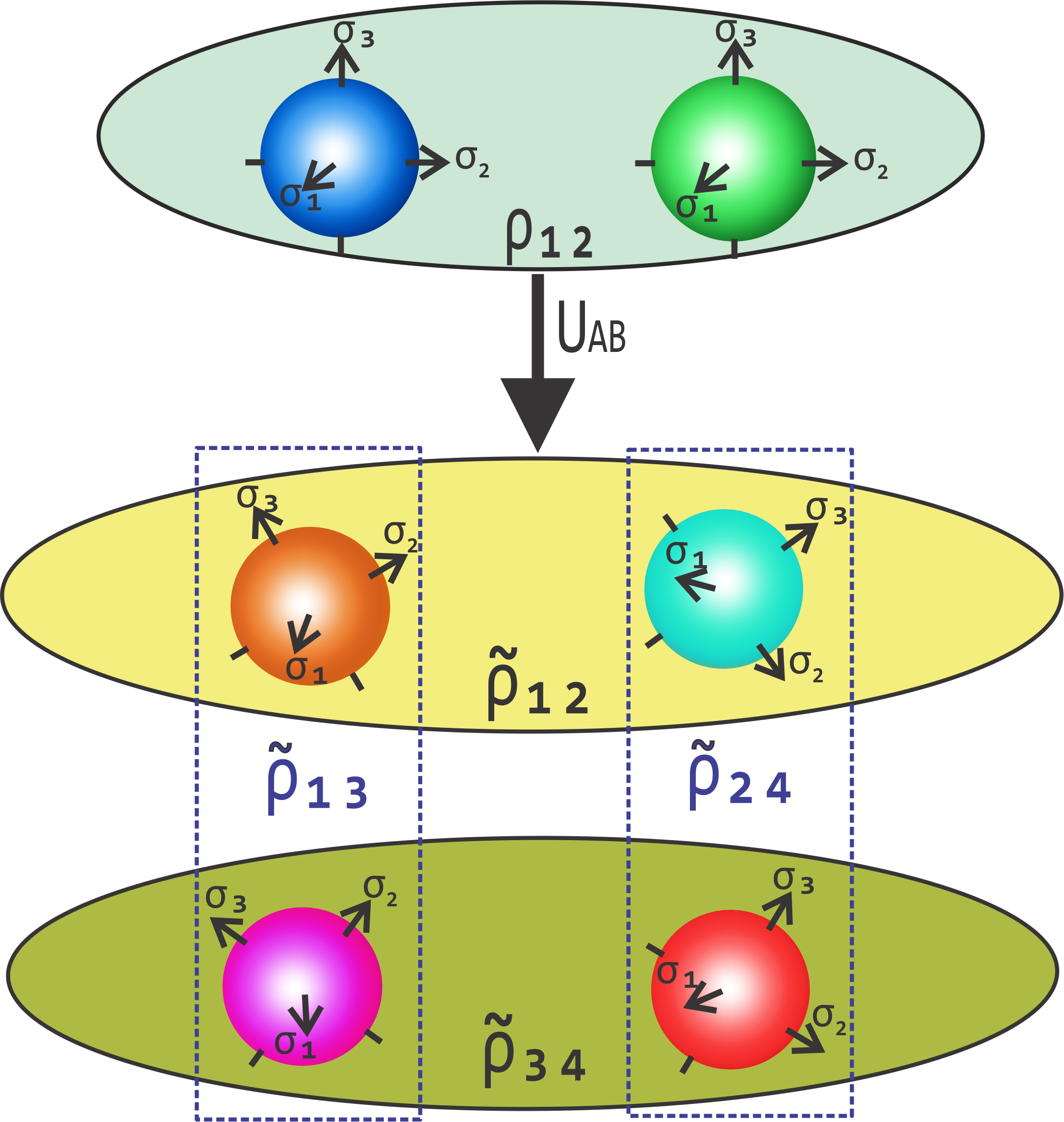}
                \caption{Nonlocal broadcasting. Only the horizontal nonlocal output pairs ($\tilde{\rho}_{12}$ and $\tilde{\rho}_{34}$) have been highlighted for clarity.}
                \label{fig:1to2nloc}
        \end{subfigure}%
        \caption{The figures depict broadcasting of entanglement. The boxes with dotted boundary highlight the local output pairs - $\tilde{\rho}_{13}$ and $\tilde{\rho}_{24}$ \cite{foot-color}.}\label{fig:1to2}
\end{figure}
\section{ $1 \rightarrow 2$ broadcasting of entanglement using asymmetric cloners}
\label{sec:1to2broadent}
\subsection{Broadcasting via local cloning}
In this section we describe the broadcasting of entanglement using asymmetric cloners where our resource state is a two qubit mixed state. The state is shared between two parties A and B with qubits numbered $1$ and $2$ respectively (as in Eq.~\eqref{eq:mix}). Both of them apply the optimal universal asymmetric Pauli cloning machine $\left(U^l_a\right)$ as in Eq.\eqref{eq:local_cloner}. The process is illustrated in Fig~\ref{fig:1to2loc}. 

After cloning and tracing out the ancillary qubits on each side, we obtain the composite system $\tilde{\rho}_{1234}$,
\begin{equation}
\tilde{\rho}_{1234}=Tr_{56}[U_A\otimes U_B (\rho_{12}\otimes \sigma_{34} \otimes \mathcal{M}_{56})U_B^{\dagger}\otimes U_A^{\dagger}],
\end{equation}
where $U_A=U_B=\left(U^l_a\right)$,$\sigma_{34}=\dyad{00}$ represents the initial blank state, and $\mathcal{M}_{56}=\dyad{00}$ represents the initial state of the ancillary qubits. $U_A(U_B)$ acts on odd(even) numbered parties.

We trace out the qubits $(2,4)$ and $(1,3)$ to obtain the local output states $\tilde{\rho}_{13}$ on A's side and $\tilde{\rho}_{24}$ on B's side respectively. The canonical expressions for the reduced density operators of local output states are as follows :
\begin{equation}
\label{eq:gen_local_localoutputs}
\begin{split}
\tilde{\rho}_{13} = \left\{p\mu\vec{x}, q\mu\vec{x}, pq\mu\mathbb{I}_3 \right\},\: \tilde{\rho}_{24} = \left\{p\mu\vec{y},q\mu\vec{y},pq\mu\mathbb{I}_3 \right\},
\end{split}
\end{equation}
where $p$ and $q$ are the parameters of the asymmetric cloning machine as defined in Eq.\eqref{eq:local_cloner}, and $\mu=\frac{1}{1-pq}$. Here $\mathbb{I}_3$ refers to the 3-dimensional identity matrix.

Next, we apply \textbf{PH} criterion (as described before in Sec.~\ref{sec:ph}) to investigate the separability of these local output states. We obtain a range involving input state parameters and the asymmetry parameter of the cloner,
\begin{eqnarray}
0\leq\|\vec{x}\|^2\leq1-4p^2q^2 \: \: \text{and}\:\:
%\text{and}\:\:\:\:\:\:\:\:\:\:\:\:\:\:\:\:\:\:\:\:\:\:\:\:\:\:\:\:\:\:\:\:\:\:\:\:\:\:\:\nonumber\\
0\leq\|\vec{y}\|^2\leq1-4p^2q^2 .
\label{eq:range_local_separability}
\end{eqnarray}
Here $\|\vec{a}\|^2=\Tr{a^\dagger a}$, and $a^\dagger$ refers to the Hermitian conjugate of $a$. 

Similarly, after tracing out appropriate qubits from the composite system, we obtain the two plausible groups of nonlocal output states. The first group consists of diagonal pairs $\tilde{\rho}_{14}$ and $\tilde{\rho}_{23}$. The second group consists of horizontal pairs $\tilde{\rho}_{12}$ and $\tilde{\rho}_{34}$.
The expressions for these groups of reduced density operators are as follows :
\begin{flalign}
&\tilde{\rho}_{14} = \left\{p\mu\vec{x}, q\mu\vec{y}, pq\mu^2\mathbb{T} \right\},
\tilde{\rho}_{23} = \left\{q\mu\vec{x}, p\mu\vec{y}, pq\mu^2\mathbb{T} \right\},\nonumber\\
\,\text{Or, }&\tilde{\rho}_{12} =\left\{p\mu\vec{x}, p\mu\vec{y}, p^2 \mu^2\mathbb{T} \right\},
\tilde{\rho}_{34} =\left\{q\mu\vec{x}, q \mu \vec{y}, q^2\mu^2\mathbb{T} \right\},
\label{eq:gen_local_nonlocaloutputs}
\end{flalign}
where $\mu=\frac{1}{1-pq}$, $\mathbb{T}$ is the correlation matrix of the initial input state $\rho_{12}$.

We again apply the \textbf{PH} criterion to determine the condition for which these pairs will be inseparable, using the determinants of $W^{l}_2$, $W^{l}_3$ and $W^{l}_4$ of both the matrices as defined in Eq.\eqref{eq:w3w4}. 

Now combining these two ranges determining the separability of the local states and inseparability of the nonlocal states, we obtain the range for optimal broadcasting of entanglement. We observe that all the above expressions and ranges reduce to the one in \cite{sourav} for the case of perfectly symmetric cloner i.e. $p=q=1/2$. 

We demonstrate the variation of the broadcasting range with the asymmetry of the cloner using a class of states called Maximally Entangled Mixed States (MEMS) as a resource. This class is chosen because it is parameterised by the amount of entanglement present in the state. This helps us in analysing the amount of concurrence needed in the input state to perform broadcasting successfully, for a given asymmetric cloner. Also, this class of states has been experimentally created by Peters {\it et al} \cite{peters}.

\begin{center}\noindent\textit{Example : Maximally Entangled Mixed States(MEMS)}\end{center}

Maximally entangled mixed states are those which possess maximal amount of entanglement for a given degree of mixedness. There are various choices for measuring both these quantities which lead to different forms of density matrix for MEMS \cite{wei}. Here, we focus on a parameterisation based on \textit{concurrence squared} as a measure of entanglement and \textit{linear entropy} as a measure of mixedness. Concurrence is defined as $C(\rho)=[max\{0,\lambda_1-\lambda_2-\lambda_3-\lambda_4\}]$. Here, $\lambda_i$ refer to the square root of the eigenvalues of the matrix $\rho(\sigma_2 \otimes \sigma_2)\rho^*(\sigma_2 \otimes \sigma_2)$, in non-increasing order of magnitude, and, $\sigma_2=\begin{pmatrix}
0 & -i \\
i & 0\\
\end{pmatrix}
$. Linear entropy is defined as $S_L(\rho)=\frac{4}{3}[1-Tr(\rho^2)]$. The density matrix for this parameterisation occurs in following two sub-classes \cite{19munro}
\begin{equation}
\begin{split}
& \rho_{MI} = \frac{r}{2}\begin{pmatrix}
1 & 0 & 0 & 1 \\
0 & \frac{2(1-r)}{r} & 0 & 0 \\
0 & 0 & 0 & 0 \\
1 & 0 & 0 & 1 \\
\end{pmatrix}, \frac{2}{3} \le r \le 1,\\
& \rho_{MII} = \begin{pmatrix}
\frac{1}{3} & 0 & 0 & \frac{r}{2} \\
0 & \frac{1}{3} & 0 & 0 \\
0 & 0 & 0 & 0 \\
\frac{r}{2} & 0 & 0 & \frac{1}{3} \\
\end{pmatrix}, 0 \le r \le \frac{2}{3},\\
\end{split}
\end{equation}
where $r$ is the concurrence of the state.

After applying the local asymmetric cloner (Eq.\eqref{eq:local_cloner}) to subclass $I$ and tracing out appropriate qubits, we get the local output pairs to be,
\begin{equation}
\begin{split}
& \tilde{\rho}^{MI}_{13} = \left\{ \{0,0,p\mu(1-r)\},\{0,0,q\mu(1-r)\},pq\mu\mathbb{I} \right\}, \\
& \tilde{\rho}^{MI}_{24} = \left\{ \{0,0,p\mu(r-1)\},\{0,0,q\mu(r-1)\}, pq\mu\mathbb{I} \right\} ,
\end{split}
\end{equation}
where $\mu=\frac{1}{1-pq}$ and $\mathbb{I}$ is the identity matrix. These states are found to be always separable. 
The two possible groups of nonlocal output states obtained after tracing out are given by,
\begin{equation}
\begin{split}
& \tilde{\rho}^{MI}_{14} = \left\{ \{0,0,p(1-r)\mu\},\{0,0,q(r-1)\mu\}, pq\mu^2\mathbb{T}^I \right\}, \\
& \tilde{\rho}^{MI}_{23} =\left\{ \{0,0,q(1-r)\mu\},\{0,0,p(r-1)\mu\}, pq\mu^2\mathbb{T}^I \right\},\\ \nonumber
\end{split}
\end{equation}
\begin{equation}
\begin{split}
\text{Or, }&\tilde{\rho}^{MI}_{12} = \left\{ \{0,0,p(1-r)\mu\},\{0,0,p(r-1)\mu\}, p^2\mu^2\mathbb{T}^I \right\}, \\
& \tilde{\rho}^{MI}_{34} =\left\{ \{0,0,q(1-r)\mu\},\{0,0,q(r-1)\mu\}, q^2\mu^2\mathbb{T}^I \right\}, \\
\end{split}
\end{equation}
where $\mu=\frac{1}{1-pq}$, $\mathbb{T}^I=\textit{diag}\{r,-r,2r-1\}$ is the correlation matrix of the initial input state $\rho_{MI}$ and $\textit{diag}$ refers to a diagonal matrix.

Employing the \textbf{PH} criterion, we find that the diagonal group of nonlocal states i.e. $\tilde{\rho}^{MI}_{14}$ and $\tilde{\rho}^{MI}_{23}$ are inseparable if the concurrence in the initial input state is greater than a function of the asymmetric parameter of the cloner given by,
\begin{equation}
\begin{split}
& r > \Bigl(\frac{\lambda (\lambda (3-2 \lambda)-4)+1}{(\lambda-1)^2}+\\
& \frac{\sqrt{(\lambda (\lambda-1)+1) (\lambda (\lambda (\lambda (5 \lambda-11)+16)-7)+1)}}{(\lambda-1)^2}\Bigr)
\end{split}
\end{equation}
where $\lambda=pq$. 
The minimum of the above function is obtained for $p=q=1/2$ (symmetric cloner), for which the corresponding concurrence is $\frac{5}{36} \left(2+\sqrt{13}\right) \approx 0.78$.

Again, employing the \textbf{PH} criterion, we find that the horizontal group of nonlocal states i.e. $\tilde{\rho}^{MI}_{12}$ and $\tilde{\rho}^{MI}_{34}$ are inseparable if the concurrence in the initial state is greater than a function of the asymmetric parameter of the cloner given by,
\begin{equation}
    r>\frac{\left(p^2+1\right)q^2}{2 (p q-1)^2}\left(\frac{\sqrt{p^4-2 p^3+4 p^2-2 p+1}}{p}+1\right)
\end{equation}

Since the local output states are always separable, the final broadcasting range coincides with the range on input and asymmetry parameters for which the nonlocal outputs are inseparable. 

We repeat the procedure for subclass $II$ for which the concurrence lies in the range $0\le r \le \frac{2}{3}$. In this case, we find that again the local outputs are always separable. However, the non-local outputs are also separable and thus this subclass is not broadcastable using local cloners. This is consistent with the fact that the initial state of this subclass has low amount of entanglement (concurrence) as compared to the previous subclass. 

\begin{figure}[h]
\centering
    \includegraphics[scale=0.8]{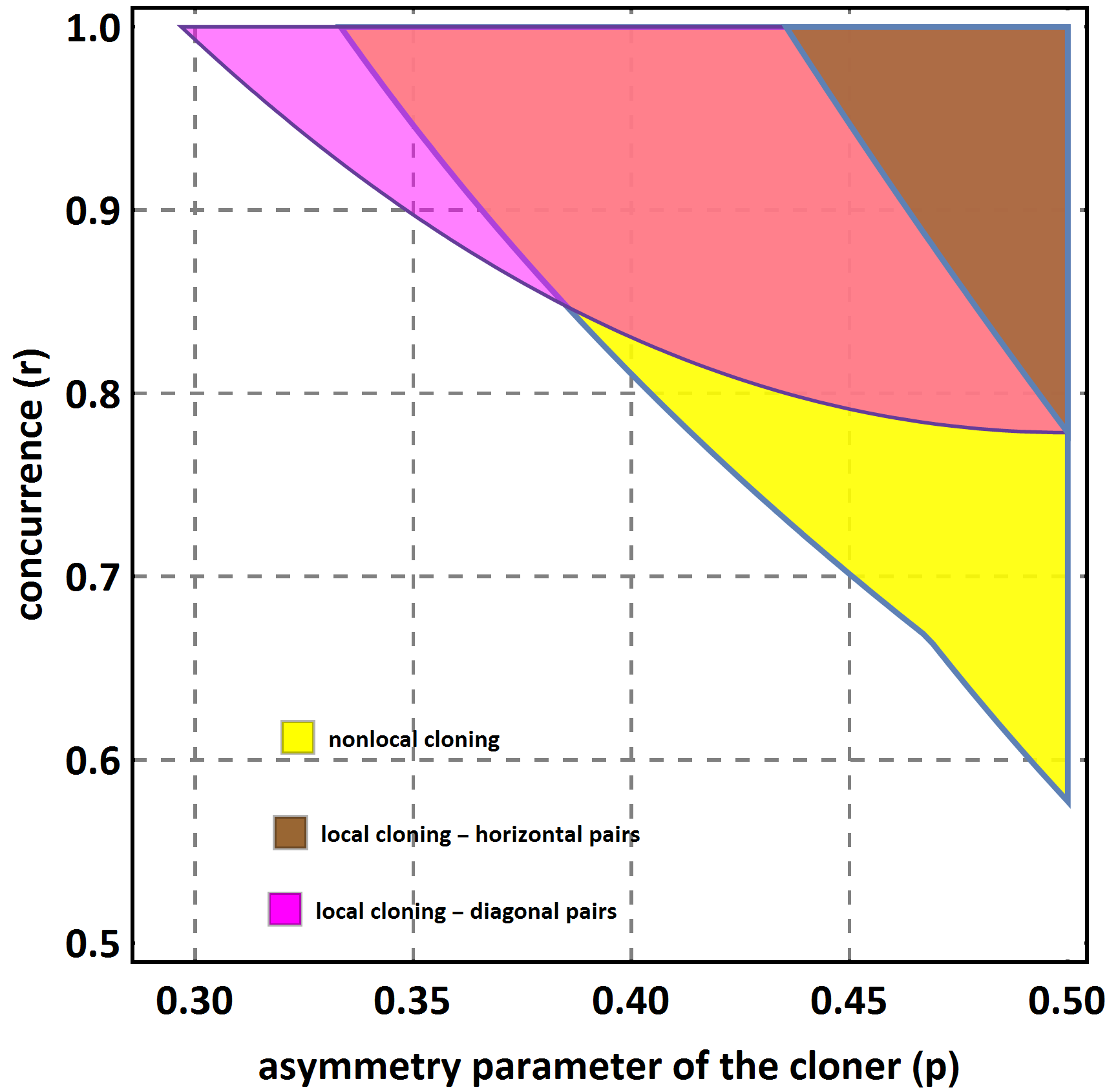}
  \caption{\scriptsize The figure shows broadcastable region for Maximally Entangled Mixed States (MEMS). The brown (dark gray) region corresponds to the case for local broadcasting when horizontal nonlocal pairs are used, whereas the magenta (medium gray) region corresponds to the use of diagonal nonlocal pairs. The yellow (light gray) region represents nonlocal broadcasting using asymmetric cloners. The horizontal axis demonstrates the variation in asymmetry of the cloner with $p=1/2$ representing a completely symmetric cloner. The vertical axis represents the concurrence of the input state. }
\label{fig:MEMS}
\end{figure}

In Fig (\ref{fig:MEMS}) we depict the broadcastable region when the input state is parameterised by the amount of concurrence $r$, and cloners with asymmetry parameter $p$ are used. The figure is symmetric about the line $p=1/2$ and hence only one half of the plot has been shown. We observe from the figure that as we increase the value of the parameter $p$ from $0$ (perfectly asymmetric) to $1/2$ (symmetric), broadcastable region in the state space increases. In the case of MEMS, we find that any cloner having $p$ less than $\sim 0.3$ is unable to broadcast entanglement in the case of MEMS when considering diagonal pairs represented by magenta (medium gray) region; and $p$ less than $\sim 0.44$ considering horizontal pairs indicated by brown (dark gray) region. This gives the impression that beyond a certain degree of asymmetry in the cloner, broadcasting becomes impossible. This observation can be understood from the intuition that the asymmetry parameter induces a trade-off in the distribution of entanglement among the two nonlocal output pairs. We also observe that the minimum concurrence needed in the input state for broadcasting to be successful using both horizontal and diagonal pairs is $\sim 0.78$. We notice that as the concurrence of input state increases, the broadcastable region increases. 

The two choices of nonlocal pairs, diagonal and horizontal, give two different regions for broadcasting. The diagonal pairs lead to a wider range than the horizontal pairs. This stems from the asymmetric nature of cloning on both ends of the input state. As a consequence, the first cloned outputs on each side i.e. parties $1$ and $2$, have similar fidelity. Also, the second clones on each side, parties $3$ and $4$, have similar fidelity (refer Fig.\ref{fig:1to2loc}). Therefore the amount of entanglement in the horizontal pairs $\tilde{\rho}_{12}$ and $\tilde{\rho}_{34}$ differs significantly. On the other hand $\tilde{\rho}_{14}$ and $\tilde{\rho}_{23}$ tend to have similar amount of entanglement, because these pairs comprise of qubits from both first and second clones. This leads to the diagonal pairs being simultaneously entangled for a larger range of values of input state parameter and the asymmetry parameter of the cloner (Fig.\ref{fig:MEMS}).

\subsection{Broadcasting via nonlocal cloning}
Here, we investigate the broadcasting of entanglement via nonlocal cloning. 
We again start with two parties sharing a two-qubit mixed state (as in Eq.\eqref{eq:mix}) but now they together apply the nonlocal optimal universal asymmetric Pauli cloning machine as in Eq.\eqref{eq:nonlocal_cloner}. The process is illustrated in Fig~\ref{fig:1to2nloc}. 

After cloning and tracing out the ancillary qubits on each side, we obtain the composite system $\tilde{\rho}_{1234}$,
\begin{equation}
\tilde{\rho}_{1234}=Tr_{56}[U_{AB} (\rho_{12}\otimes \sigma_{34} \otimes \mathcal{M}_{56})U_{AB}^{\dagger}],
\end{equation}
where $U_{AB}=\left(U^{nl}_a\right)$,$\sigma_{34}=\dyad{00}$ represents the initial blank state, and $\mathcal{M}_{56}=\dyad{00}$ represents the initial state of the ancillary qubits. $U_{AB}$ acts on the combined state of all the six parties. 

We trace out appropriate qubits to obtain the local output states $\tilde{\rho}_{13}$ on A's side and $\tilde{\rho}_{24}$ on B's side respectively. The canonical expressions for the reduced density operators of local output states are given by :
\begin{equation}
\label{eq:gen_nonlocal_localoutputs}
\begin{split}
&\tilde{\rho}_{13} = \left\{\kappa_1\vec{x}, \kappa_2\vec{x}, \frac{\kappa_1q}{1+p}\mathbb{I}_3 \right\},\tilde{\rho}_{24} = \left\{\kappa_1\vec{y}, \kappa_2\vec{y}, \frac{\kappa_1q}{1+p}\mathbb{I}_3 \right\}.
\end{split}
\end{equation}
where $\kappa_1 = p(1+p)/(2-3pq),\kappa_2=q(2-p)/(2-3pq)$.

By applying \textbf{PH} criterion, we obtain the following ranges involving input state parameters and the parameter of asymmetry within which the local outputs, $\tilde{\rho}_{13}$ and $\tilde{\rho}_{24}$, are separable. 
\begin{flalign}
0\leq\|\vec{x}\|^2\leq\frac{1-2pq}{(1-pq)^2} \: \: \text{and}\:\: 0\leq\|\vec{y}\|^2\leq\frac{1-2pq}{(1-pq)^2} \:.
%\text{and}\:\:\:\:\:\:\:\:\:\:\:\:\:\:\:\:\:\:\:\:\:\:\:\:\:\:\:\:\:\:\:\:\:\:\:\:\:\:\:\nonumber\\
\label{eq:range_nonlocal_separability}
\end{flalign}

Similarly, after tracing out appropriate qubits from the composite system, we obtain the two plausible groups of nonlocal output states. The first group consists of diagonal pairs $\tilde{\rho}_{14}$ and $\tilde{\rho}_{23}$. The second group consists of horizontal pairs $\tilde{\rho}_{12}$ and $\tilde{\rho}_{34}$. The expressions for these groups of reduced density operators are as follows :
\begin{flalign}
& \tilde{\rho}_{14} = \left\{\kappa_1\vec{x}, \kappa_2\vec{y}, \frac{\kappa_1q}{1+p}\mathbb{T} \right\},
\tilde{\rho}_{23} = \left\{\kappa_1\vec{y}, \kappa_2\vec{x}, \frac{\kappa_1q}{1+p}\mathbb{T} \right\};\nonumber\\
& \text{Or, }\,\,\tilde{\rho}_{12} = \left\{\kappa_1\vec{x}, \kappa_1\vec{y}, \kappa_1\mathbb{T} \right\}, 
 \tilde{\rho}_{34} = \left\{\kappa_2\vec{x}, \kappa_2\vec{y}, \kappa_2\mathbb{T} \right\},
\label{eq:gen_nonlocal_nonlocaloutputs}
\end{flalign}
\noindent where $\kappa_1 = p(1+p)/(2-3pq),\kappa_2=q(2-p)/(2-3pq)$, $\mathbb{T}$ is the correlation matrix of the initial input state $\rho_{12}$.

We again apply the \textbf{PH} criterion to determine the condition for which both the states in either of these groups will be inseparable, using the determinants $W^{nl}_2$, $W^{nl}_3$ and $W^{nl}_4$ as defined in Eq.\eqref{eq:w3w4}. 

Now combining these two ranges which determine the separability of the local states and inseparability of the nonlocal states, we obtain the range for optimal broadcasting of entanglement. We observe that all the above expressions and ranges reduce to the one in \cite{sourav} for the case of perfectly symmetric cloner i.e. $p=q=1/2$.

Similar to the local cloning case, we again exemplify the variation of the broadcasting range using the class of MEMS as our resource state. 

\begin{center}\noindent\textit{Example : Maximally Entangled Mixed States(MEMS)}\end{center}
Once the nonlocal cloner (Eq.\eqref{eq:nonlocal_cloner}) is applied to the state of subclass $I$, the nonlocal outputs obtained become,
\begin{equation}
\begin{split}
& \tilde{\rho}^{MI}_{12} = \left\{ \{0,0,\kappa_1(1-r)\},\{0,0,\kappa_1(r-1)\}, \kappa_1\mathbb{T}^I \right\}, \\
& \tilde{\rho}^{MI}_{34} = \left\{ \{0,0,\kappa_2(1-r)\},\{0,0,\kappa_2(r-1)\}, \kappa_2\mathbb{T}^I \right\}, \\
\end{split}
\end{equation}
with $\kappa_1 = p(1+p)/(2-3pq),\kappa_2=q(2-p)/(2-3pq)$, and $\mathbb{T}^I=$\textit{diag}$\{r,-r,2r-1\}$ being the correlation matrix of the initial input state $\rho_{MI}$. Here \textit{diag} refers to a diagonal matrix.\\
The inseparability range of these output states are then given by :
\begin{equation}
\begin{split}
&(p (p+1) \left(p^2 (r+1)+p \left(K+r-2\right)-K+1\right)>0 \\
& \text{ \& } 3 p>1) \text{ Or, } 22 p+\sqrt{502+318 \sqrt{5}}>25+9 \sqrt{5},
\end{split}
\end{equation}
where $K = \sqrt{4 p^2-2 p+2}$.\\
We omit the second possible group of diagonal nonlocal states i.e. $\tilde{\rho}^{MI}_{14}$ and $\tilde{\rho}^{MI}_{23}$ because they are found to be separable. We hypothesize that this is true always, even for the general two-qubit mixed input state. More particularly, the hypothesis is that the diagonal pair of nonlocal output states are always separable in the case of nonlocal broadcasting. 
Similarly, the local output states are described as follows:
\begin{equation}
\begin{split}
& \tilde{\rho}^{MI}_{13} = \left\{ \{0,0,\kappa_1(1-r)\},\{0,0,\kappa_2(1-r)\}, \frac{pq}{2-3pq}\mathbb{I} \right\},\\
& \tilde{\rho}^{MI}_{24} = \left\{ \{0,0,\kappa_1(r-1)\},\{0,0,\kappa_2(r-1)\}, \frac{pq}{2-3pq}\mathbb{I} \right\}, \\
\end{split}
\end{equation}
where $\kappa_1 = p(1+p)/(2-3pq)$, and $\kappa_2=q(2-p)/(2-3pq)$.
The local outputs are found to be unconditionally separable. Thus, the final range for broadcasting for subclass $I$ is the same as the inseparability range.

We repeat the procedure for subclass $II$, where the nonlocal outputs obtained are
\begin{equation}
\begin{split}
& \tilde{\rho}^{MII}_{12} = \left\{ \{0,0,\frac{\kappa_1}{3}\},\{0,0,\frac{-\kappa_1}{3}\}, \kappa_1\mathbb{T}^{II} \right\}, \\
& \tilde{\rho}^{MII}_{34} = \left\{ \{0,0,\frac{\kappa_2}{3}\},\{0,0,\frac{-\kappa_2}{3}\}, \kappa_2\mathbb{T}^{II} \right\}, \\
\end{split}
\end{equation}
with $\kappa_1 = p(1+p)/(2-3pq),\kappa_2=q(2-p)/(2-3pq)$, and $\mathbb{T}^{II}=$\textit{diag}$\{r,-r,1/3\}$ being the correlation matrix of the initial input state $\rho_{MII}$. Here \textit{diag} refers to a diagonal matrix.

These output states are inseparable as long as concurrence in the initial state is larger than a value given by :
\begin{equation}
r > \frac{1}{\sqrt{3}}\:\cross\:max\biggl(\sqrt{\frac{q^2 (p(5p-4)+3)}{p^2 (p+1)^2}}, \frac{p \sqrt{p (5p-6)+4}}{\sqrt{3} | q(2-p)|}\biggr)
\end{equation}
Here also, we omit the second possible group of diagonal nonlocal states i.e. $\tilde{\rho}^{MII}_{14}$ and $\tilde{\rho}^{MII}_{23}$ because they are found to be separable.
Similarly, the local output states are described as follows:
\begin{equation}
\begin{split}
& \tilde{\rho}^{MII}_{13} = \left\{ \{0,0,\frac{\kappa_1}{3}\},\{0,0,\frac{\kappa_2}{3}\}, \frac{\kappa_1q}{1+p}\mathbb{I} \right\}, \\
& \tilde{\rho}^{MII}_{24} = \left\{ \{0,0,\frac{-\kappa_1}{3}\},\{0,0,\frac{-\kappa_2}{3}\}, \frac{\kappa_1q}{1+p}\mathbb{I} \right\}. \\
\end{split}
\end{equation}
where $\kappa_1 = p(1+p)/(2-3pq)$, and $\kappa_2=q(2-p)/(2-3pq)$. 
The local outputs are found to be unconditionally separable. Thus, the final range for broadcasting for subclass $II$ is the same as its inseparability range.

In Fig (\ref{fig:MEMS}), the yellow (light gray) region depicts the broadcastable zone for the nonlocal case. All the observations with respect to the variation of broadcasting range from the local case hold true for the nonlocal case as well. The minimum concurrence needed in the input state for broadcasting to be successful using nonlocal cloner is $\sim 0.58$, which is smaller than the concurrence needed in case of broadcasting with local operations. It reestablishes the fact that nonlocal cloners perform better than the local ones.

\section{ $1 \rightarrow 2$ broadcasting of correlations beyond entanglement using asymmetric cloners}
\label{sec:1to2broaddis}

Recent researches reveal the presence of quantum correlations beyond entanglement (QCsbE) \cite{discord1, discord2, discordresource}. Entanglement measures are insufficient to quantify all such correlations \cite{discord1,discord2,discord3,discord4}. In this section we investigate the optimal broadcasting of QCsbE using universal optimal asymmetric cloners. We have chosen geometric discord ($D_G$, given by Eq.~\ref{eq:gdiscord}) as our measure to quantify the degree of QCsbE in the states.

\subsection{Via local cloning}

We again start with the general mixed state (refer Eq.~\ref{eq:mix}) and apply local cloning(Eq.\eqref{eq:local_cloner}) on each side. After tracing out qubits we obtain the local output states - $\tilde\rho_{13}$ and $\tilde\rho_{24}$ and the two possible groups of nonlocal output states $\tilde\rho_{14}$ and $\tilde\rho_{23}$, or, $\tilde\rho_{12}$ and $\tilde\rho_{34}$.
To achieve optimal broadcasting we require the nonlocal states to possess non-zero discord whereas the local pairs should possess zero discord i.e. 
\begin{equation}
\begin{split}
& \{(D_{G}(\tilde\rho_{14})>0, D_{G}(\tilde\rho_{23})>0)  \\
 \text{ or } &(D_{G}(\tilde\rho_{12})>0, D_{G}(\tilde\rho_{34})>0)\}  \\
 \text{ and }& D_{G}(\tilde\rho_{24}) = 0 \text{ and }  D_{G}(\tilde\rho_{13}) = 0.
\end{split}
\label{eq:optimalbroaddisc}
\end{equation}

\noindent \begin{theorem}
\textit{Given a two qubit general mixed state $\rho_{12}$ and optimal universal asymmetric Pauli cloning transformation ($U^l_a$), it is impossible to broadcast the QCsbE optimally within $\rho_{12}$ into two lesser quantum correlated states: \{$\tilde{\rho}_{14}$, $\tilde{\rho}_{23}$\} or \{$\tilde{\rho}_{12}$, $\tilde{\rho}_{34}$\}.  
}
\end{theorem}
\noindent \emph{Proof:}
A two qubit general mixed state $\rho_{12}$ (Eq.~\ref{eq:mix}) is taken and optimal universal asymmetric Pauli cloner (Eq.\eqref{eq:local_cloner}) is applied locally at both ends. The local output states are obtained as in Eq.~\eqref{eq:gen_local_localoutputs} and the nonlocal output states are obtained as in Eq.~\eqref{eq:gen_local_nonlocaloutputs}.
%start here
We calculate the geometric discord for the local output states using Eq.~\eqref{eq:gdiscord} and the expressions are:
\begin{equation}
\begin{split}
& D_{G}(\tilde\rho_{13}) = D_{G}(\tilde\rho_{24}) = \frac{p^2q^2}{2 \left(1-pq\right)^2}
\end{split}
\end{equation}
where $p+q=1$.

When we impose the condition for optimal broadcasting i.e. $D_{G}(\tilde\rho_{13})=0 \text{ and } D_{G}(\tilde\rho_{24})=0$, we find that for this to happen either $p=0$ or $q=0$. The former condition, i.e. $p=0$ leads to $D_{G}(\tilde\rho_{14})=D_{G}(\tilde\rho_{23})=D_{G}(\tilde\rho_{12})=0$, whereas the latter condition leads to $D_{G}(\tilde\rho_{14})=D_{G}(\tilde\rho_{23})=D_{G}(\tilde\rho_{34})=0$. Therefore, it is impossible to have any of the groups of nonlocal pairs to be correlated simultaneously, subject to these constraints. This indicates that given a two qubit mixed state, $\rho_{12}$, it is impossible to optimally broadcast QCsbE to obtain two lesser correlated states using universal optimal asymmetric local cloners.
\subsection{Via nonlocal cloning}
We again start with the general mixed state (Eq.~\ref{eq:mix}) and apply nonlocal cloning (Eq.\eqref{eq:nonlocal_cloner}) on the qubits. The condition for broadcasting is the same as Eq.\eqref{eq:optimalbroaddisc}. 
\noindent \begin{theorem}
\label{theo:nonlocal_broad}
\textit{Given a two qubit general mixed state $\rho_{12}$ and optimal universal nonlocal asymmetric Pauli cloning transformation ($U^{nl}_a$), it is impossible to broadcast the QCsbE optimally within $\rho_{12}$ into two lesser quantum correlated states: \{$\tilde{\rho}_{14}$, $\tilde{\rho}_{23}$\} or \{$\tilde{\rho}_{12}$, $\tilde{\rho}_{34}$\}.  
}
\end{theorem}
\noindent \emph{Proof:}
A two qubit general mixed state $\rho_{12}$ (Eq.~\eqref{eq:mix}) is taken and an optimal universal asymmetric Pauli cloner is applied nonlocally (Eq.\eqref{eq:nonlocal_cloner}). The local output states are obtained as in Eq.~\eqref{eq:gen_nonlocal_localoutputs} and the nonlocal output states are obtained as in Eq.~\eqref{eq:gen_nonlocal_nonlocaloutputs}. We calculate the discord for the local output states using Eq.~\eqref{eq:gdiscord} and the expressions are:
\begin{equation}
\begin{split}
& D_{G}(\tilde\rho_{13}) = D_{G}(\tilde\rho_{24}) = \frac{p^2q^2}{2 \left(2-3pq\right)^2}
\end{split}
\end{equation}
where $p+q=1$.

When we impose the condition for optimal broadcasting i.e. $D_{G}(\tilde\rho_{13})=0 \text{ and } D_{G}(\tilde\rho_{24})=0$, we find that for this to happen either $p=0$ or $q=0$. The former condition, i.e. $p=0$ leads to $D_{G}(\tilde\rho_{14})=D_{G}(\tilde\rho_{23})=D_{G}(\tilde\rho_{12})=0$, whereas the latter condition leads to $D_{G}(\tilde\rho_{14})=D_{G}(\tilde\rho_{23})=D_{G}(\tilde\rho_{34})=0$. In both these cases, it is impossible to have non-zero discord in both the nonlocal pairs simultaneously. This impossibility holds true whether horizontal or diagonal groups are considered as the nonlocal outputs. Hence, we arrive at the conclusion that given a two qubit mixed state, $\rho_{12}$, it is impossible to optimally broadcast QCsbE to obtain two lesser correlated states even using universal optimal asymmetric nonlocal cloners.
\section{ $1 \rightarrow 3$ broadcasting of entanglement via successive use of $1 \rightarrow 2$ asymmetric cloners}
\label{sec:1to3broadentsucc}
In broadcasting, generating only two entangled states from a given entangled state is not always the limit. If there is exigency of creating greater number of weaker entangled states starting from a single entangled state, then there are different possibilities to explore. In this section, we explore the idea of generating three entangled pairs with the successive use of $1 \rightarrow 2$ asymmetric cloners. We also refer to this process as successive broadcasting. For this study, we use non-maximally entangled state as our resource state, $\rho_{12}=\dyad{\psi_{12}}$, where,
\begin{equation}
\ket{\psi_{12}} = \sqrt{k}\ket{00} + \sqrt{1-k}\ket{11}, \: \: 0 \le k \le 1.
\label{eq:nme}
\end{equation}
\subsection{Broadcasting via local cloning}
\begin{figure}[h]
\centering
    \includegraphics[scale=0.3]{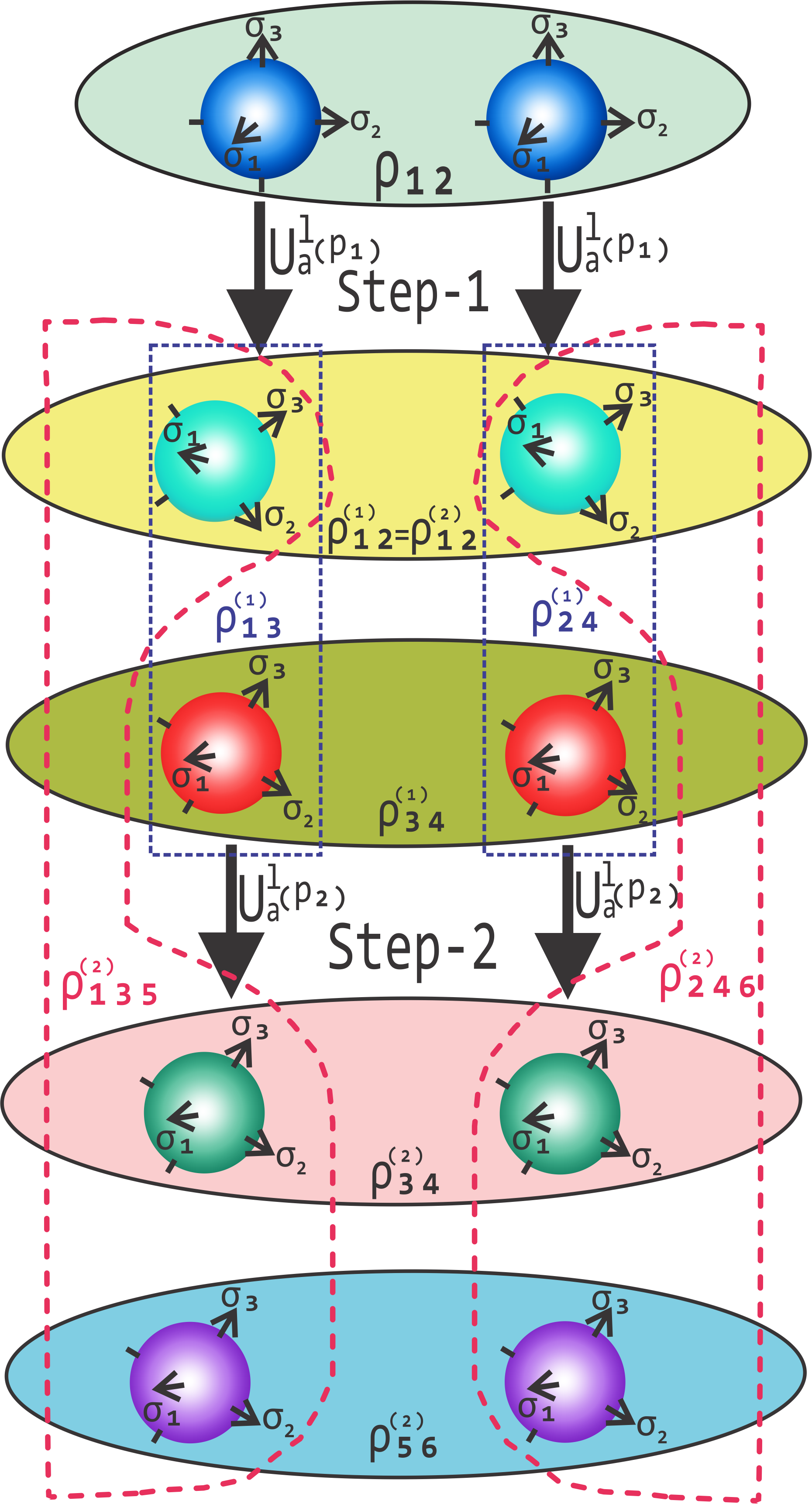} 
  \caption{\noindent \scriptsize The figure shows the two step procedure to carry out $1 \rightarrow 3$ broadcasting by successive application of asymmetric cloners locally. In step-$1$ the cloner $U^l_a$ parameterised by $p_1$ is applied on both sides. In step-$2$ the cloner $U^l_a$ parameterised by $p_2$ is applied on both sides of $\rho^{(1)}_{34}$, which is a nonlocal pair obtained after first step. The outputs in first step are super-scripted with $(1)$ while the final output states in second step are super-scripted by $(2)$. Here only the horizontal nonlocal pairs have been highlighted for clarity. The dark blue dotted box comprises of the local output pairs obtained after application of first cloner while dashed envelope contains those after application of the second cloner \cite{foot-color}.}
\label{fig:1to3sL}
\end{figure}

The method involves creation of two non-local pairs via regular $1 \rightarrow 2$ local broadcasting initially. This is followed by the use one of the nonlocal pairs again to generate more entangled pairs. It is a two-step procedure as demonstrated in Fig (\ref{fig:1to3sL}). In step $1$, we apply $U^l_a$(Eq \eqref{eq:local_cloner}) on either side of $\rho_{12}$, with asymmetry parameters of the cloner being labelled as $\{p_1,q_1\}$, where $p_1 + q_1 =1$. As a consequence of this we obtain local pairs given by $\rho^{(1)}_{kl}$, where $\{k,l\} \in \{\{1,3\},\{2,4\}\}$. The two groups of nonlocal pairs obtained are - diagonal $(\rho^{(1)}_{14} \text{ and } \rho^{(1)}_{23})$ and horizontal $(\rho^{(1)}_{12} \text{ and } \rho^{(1)}_{34})$. For the second iteration, we need to choose a nonlocal state with higher amount of entanglement. The members of the former group (diagonal pairs) have the same amount of entanglement (measured by concurrence), and hence not good choice for the purpose. In the latter group (horizontal pairs), $\rho^{(1)}_{34}$ has higher concurrence than $\rho^{(1)}_{12}$ for $0<p_1<1/2$ and vice-versa holds true for $1/2<p_1<1$. Thus, there exists a symmetry about $p_1=1/2$. For this reason, in step $2$, we choose $\rho^{(1)}_{34}$ without loss of generality. We apply $U^l_a$(Eq \eqref{eq:local_cloner}) with different set of asymmetry parameters, i.e. $\{p_2,q_2\}$, where $p_2 + q_2 =1$ on both sides of $\rho^{(1)}_{34}$. Finally we trace out appropriate qubits to obtain the nonlocal pairs represented by $\rho^{(2)}_{ij}$, where $i \in \{1,3,5\},j \in \{2,4,6\}$. The final local pairs obtained are given by $\rho^{(2)}_{kl}$, where $\{k,l\} \in \{\{1,3\},\{1,5\},\{3,5\},\{2,4\},\{4,6\},\{2,6\}\}$.
\noindent \begin{theorem}
\label{theo:broadloc1to3s}
\textit{Given a non-maximally entangled state $\ket{\psi_{12}}$, it is impossible to broadcast entanglement (optimally or non-optimally) into three lesser entangled states via successive application of two different (in terms of asymmetry parameter) optimal universal asymmetric Pauli cloning transformations ($U^{l}_a$).}
\end{theorem}
\noindent \emph{Proof:}
There are six possible groups of non-local pairs that can be obtained at the end of both the steps described above. One such group of non-local states in which all pairs are horizontal, is as follows :
\begin{equation}
\begin{split}
& \rho^{(2)}_{12} = \biggl\{ \{0,0,KP_1\},\{0,0,KP_1\},P_1^2\mathbb{T}^{N} \biggr\}, \\
& \rho^{(2)}_{34} = \biggl\{ \{0,0,KQ_1P_2\},\{0,0,KQ_1P_2\},(Q_1P_2)^2\mathbb{T}^{N} \biggr\}, \\
& \rho^{(2)}_{56} = \biggl\{ \{0,0,KQ_1Q_2\},\{0,0,KQ_1Q_2\},(Q_1Q_2)^2\mathbb{T}^{N} \biggr\}, \\
\end{split}
\end{equation}
where $K=(2k-1),Q_i=q_i/(q_i + p_i^2),P_i=p_i/(q_i + p_i^2),\mathbb{T}^{N}=\textit{diag}\{2\sqrt{(1-k)k},-2\sqrt{(1-k)k},1\}$, is the correlation matrix of the initial input state $\rho_{12}=\dyad{\psi_{12}}$. Here \textit{diag} represents a diagonal matrix.

By employing the \textbf{PH} criterion, we find that it is impossible to have all the pairs inseparable simultaneously. The procedure is repeated for all other groups of non-local pairs and the inference is the same in all the cases. 
Hence, we arrive at a conclusion that given a non-maximally entangled state $\ket{\psi_{12}}$, it is impossible to broadcast entanglement (optimally or non-optimally) into three lesser entangled states, via successive application of two optimal universal asymmetric Pauli cloning transformations ($U^{l}_a$ parameterised by $(p_1,q_1)$ and $(p_2,q_2)$) in two successive steps. 
\subsection{Broadcasting via nonlocal cloning}
\begin{figure}[t]
\centering
    \includegraphics[scale=0.3]{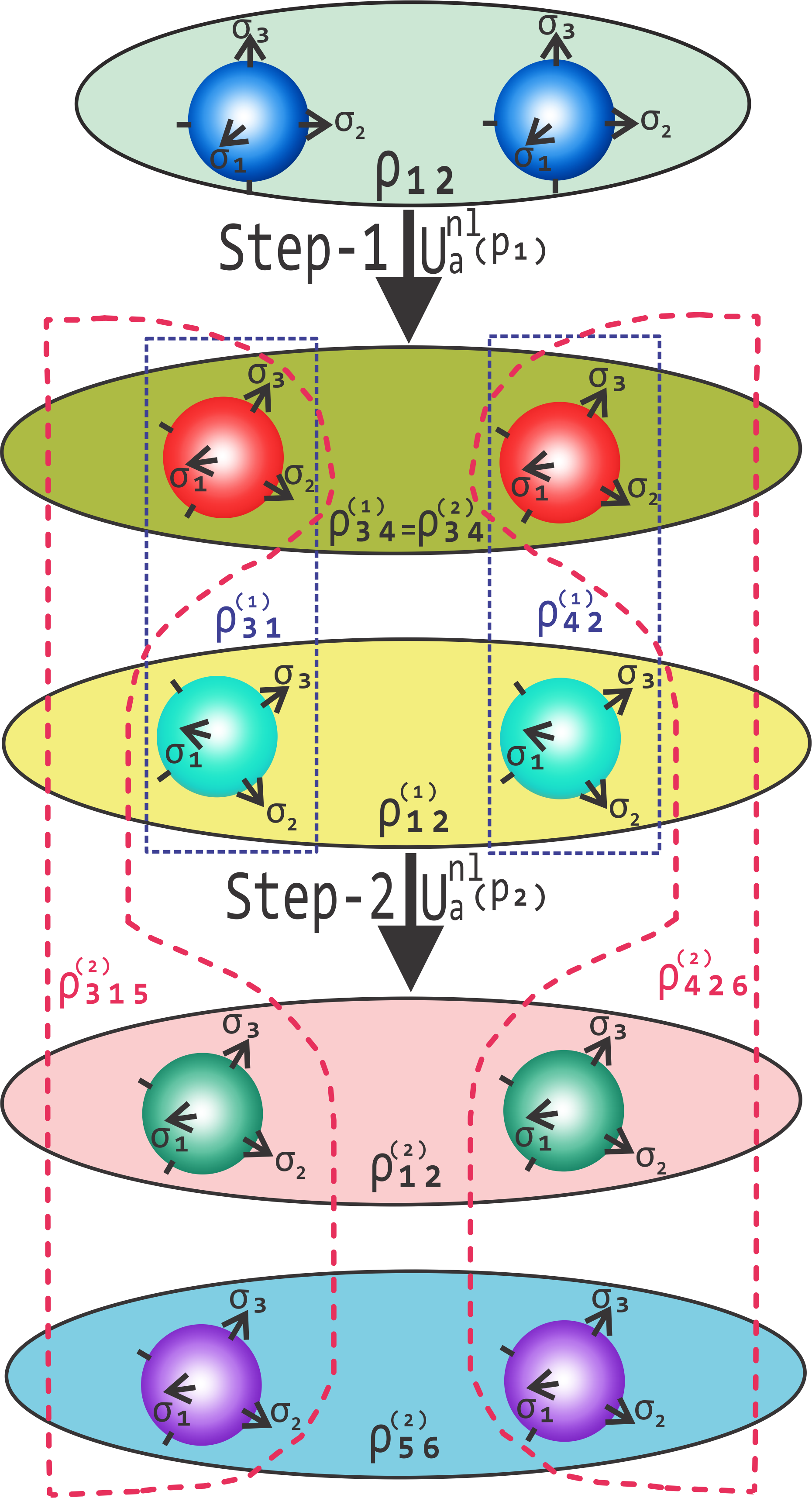}
  \caption{\noindent \scriptsize The figure depicts the two step procedure to carry out $1 \rightarrow 3$ broadcasting by successive application of asymmetric cloners nonlocally on the input resource. In step-$1$ the cloner $U^{nl}_a$ parameterised by $p_1$ is applied to the input state. In step-$2$ the cloner $U^{nl}_a$ parameterised by $p_2$ is applied to $\rho^{(1)}_{12}$, which is a nonlocal pair obtained after first step. The intermediate output pairs after the first cloning step are super-scripted with $(1)$; while the final output states are super-scripted by $(2)$ and only the horizontal nonlocal pairs (that are inseparable) have been highlighted for clarity. The dotted boxes comprise the local output pairs obtained after each step: dark-blue box contains those from step-$1$ while dashed envelope contains those from step-$2$ \cite{foot-color}.}
\label{fig:1to3snL}
\end{figure}

Similar to the case of local cloning, this time we use nonlocal cloner successively, as demonstrated in Fig (\ref{fig:1to3snL}). Unlike the local case, we discover that it is possible to broadcast entanglement into more than two pairs using nonlocal cloners. 

In step $1$, we apply $U^{nl}_a$(Eq \eqref{eq:nonlocal_cloner}) on $\rho_{12}$, with asymmetry parameters of the cloner being $\{p_1,q_1\}$, where $p_1 + q_1 =1$. As a consequence of this we obtain nonlocal pairs given by $\rho^{(1)}_{ij}$, where $i \in \{1,3\},j \in \{2,4\}$, and local pairs given by $\rho^{(1)}_{kl}$, where $\{k,l\} \in \{\{1,3\},\{2,4\}\}$. The diagonal nonlocal pairs i.e. $\rho^{(1)}_{14}$ and $\rho^{(1)}_{23}$ were found to be separable. Among horizontal nonlocal pairs, $\rho^{(1)}_{34}$ has higher concurrence than $\rho^{(1)}_{12}$ for $0<p_1<1/2$ and vice-versa holds true for $1/2<p_1<1$. There is symmetry about $p_1=1/2$. In step $2$, without loss of generality, we choose $\rho^{(1)}_{12}$ and apply $U^{nl}_a$(Eq \eqref{eq:nonlocal_cloner}) with a different set of asymmetry parameters, i.e. $\{p_2,q_2\}$, where $p_2 + q_2 =1$. As a result, the nonlocal pairs obtained after the application of two cloners and tracing out remaining qubits are represented by $\rho^{(2)}_{ij}$, where $i \in \{1,3,5\},j \in \{2,4,6\}$, and the local pairs obtained are given by $\rho^{(2)}_{kl}$, where $\{k,l\} \in \{\{1,3\},\{1,5\},\{3,5\},\{2,4\},\{4,6\},\{2,6\}\}$.

Among all the nonlocal output pairs possible, the following horizontal pairs were found to be inseparable :
\begin{equation}
\begin{split}
& \rho^{(2)}_{12} = \biggl\{ \{0,0,\tau_1\tau_2K\},\{0,0,\tau_1\tau_2K\}, \tau_1\tau_2\mathbb{T}^{N} \biggr\}, \\
& \rho^{(2)}_{34} = \biggl\{ \{0,0,\zeta_1K\}, \{0,0,\zeta_1K\}, \zeta_1\mathbb{T}^{N} \biggr\}, \\
& \rho^{(2)}_{56} = \biggl\{ \{0,0,\tau_1\zeta_2\},\{0,0,\tau_1\zeta_2K\}, \tau_1\zeta_2\mathbb{T}^{N} \biggr\}.
\end{split}
\end{equation}

The local output pairs obtained were,
\begin{equation}
\begin{split}
& \rho^{(2)}_{13} = \rho^{(2)}_{24} = \biggl\{ \{0,0,\tau_1\tau_2K\},\{0,0,\zeta_1K\}, \frac{p_1q_1\tau_2}{\eta_1}\mathbb{I} \biggr\}, \\
& \rho^{(2)}_{35} = \rho^{(2)}_{46} = \biggl\{ \{0,0,\zeta_1K\},\{0,0,\tau_1\zeta_2K\}, \frac{p_1q_1\zeta_2}{\eta_1}\mathbb{I} \biggr\}, \\
& \rho^{(2)}_{15} = \rho^{(2)}_{26} = \biggl\{ \{0,0,\tau_1\tau_2K\},\{0,0,\tau_1\zeta_2K\}, \frac{p_2q_2}{\eta_2}\mathbb{I} \biggr\},
\end{split}
\end{equation}
where $K=(2k-1),\eta_i=(2-3p_iq_i),\zeta_i=q_i(2-p_i)/\eta_i,\tau_i=p_i(1+p_i)/\eta_i$ and $\mathbb{T}^{N}$ is the correlation matrix of the initial input state $\rho_{12}=\dyad{\psi_{12}}$.\\

\begin{figure}[t]
\centering
   \includegraphics[scale=0.5]{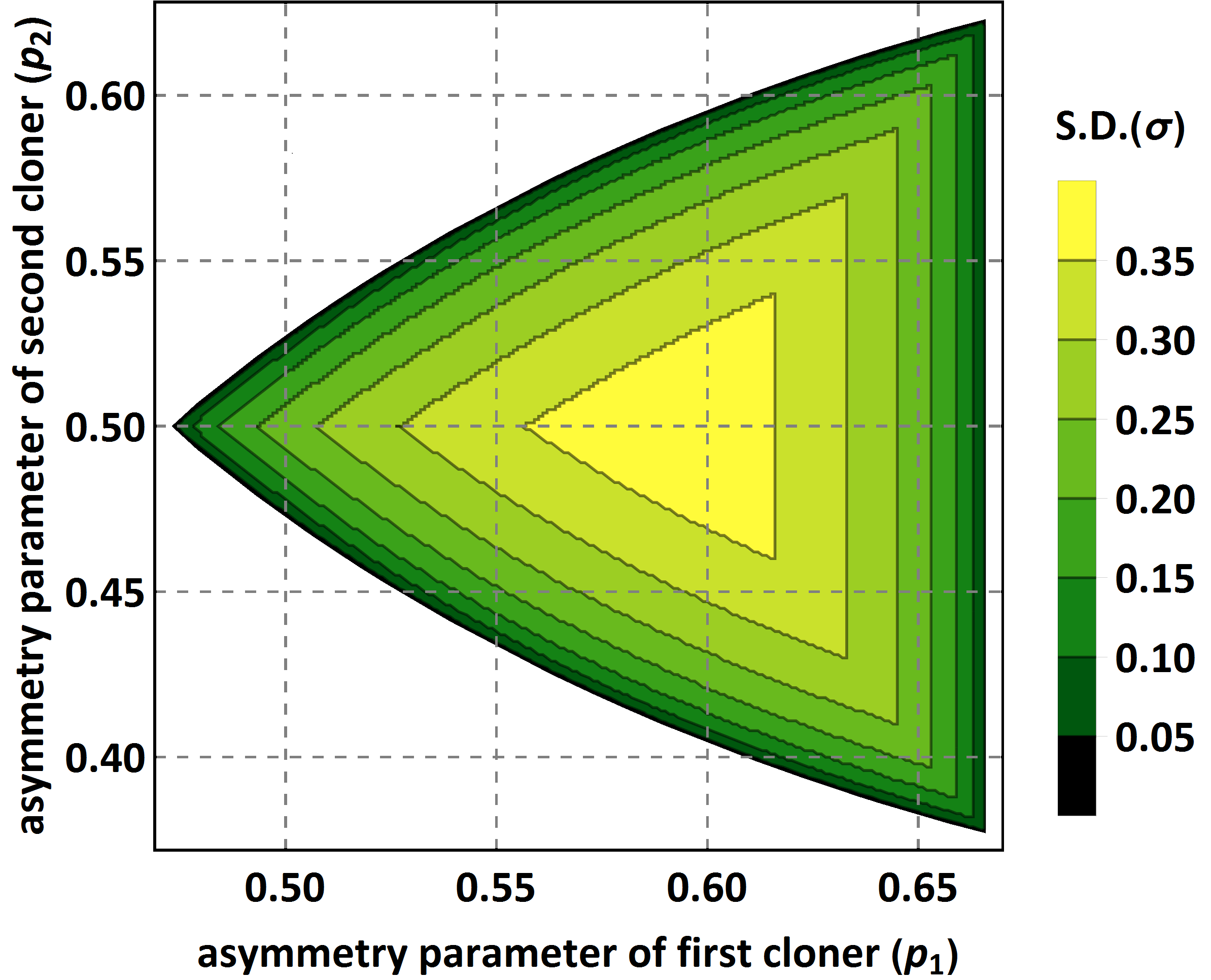} 
  \caption{\noindent \scriptsize The figure shows a contour plot of the allowed range of values of $k$ for a successful broadcasting, given the successive application of cloners parameterised by $p_1$ and $p_2$. The hue depicts the value of $\sigma$ such that the broadcasting range is $(0.5 - \sigma) \le k \le (0.5 + \sigma)$.}
\label{fig:succs1to3one}
\end{figure}

We apply the \textbf{PH} criterion to determine the condition for which these nonlocal output pairs will be inseparable, and all the local output pairs will be simultaneously separable. We obtain the broadcasting region involving three parameters : $k$ (input state parameter), $p_1$ (asymmetric parameter of first nonlocal cloner) and $p_2$ (asymmetric parameter of second nonlocal cloner). In figure (\ref{fig:succs1to3one}) we plot the variation of the broadcasting range with the asymmetry parameters of the two cloners applied. The hue chart on the right of the figure depicts the values of $\sigma$, such that the range of allowed values of $k$ is $(0.5-\sigma) \le k \le (0.5+\sigma)$. We observe that for all the input states having $0.13 \le k \le 0.87$, there exists a set of asymmetric cloners parameterised by $\{p_1,p_2\}$, which can be used to successfully broadcast the given state into three lesser entangled pairs optimally. The approximate ranges for the asymmetry parameter of the two cloners are $0.48 \le p_1 \le 0.67$ and $0.38 \le p_2 \le 0.62$. It is interesting to see that the maximum range of allowed values for $k$ is not achieved by the successive application of two symmetric cloners i.e. ($p_1 = 0.5, p_2=0.5$). The maximum range is achieved when the first cloner is slightly asymmetric i.e. $(p_1 \approx 0.6, p_2 \approx0.5)$. The reason for this is as follows : One of the nonlocal pairs produced in step $1$ is used to generate more nonlocal pairs in step $2$ of the procedure, and hence it is beneficial for that pair to have higher entanglement than the other nonlocal pair. Therefore, using a slightly asymmetric cloner in the first step is beneficial for the entire process. If one chooses the alternative nonlocal pair in the first step i.e. $\rho^{(1)}_{12}$ instead of $\rho^{(1)}_{34}$, a similar plot would be obtained in the region $0<p_1<1/2$, as the situation is completely symmetric about $p_1=1/2$. However, it is interesting to note that the plot is symmetric about $p_2=1/2$. A symmetric cloner is the best choice in the second step. Therefore, it is possible to successfully broadcast entanglement and create three entangled pairs from one by successively using $1 \rightarrow 2$ nonlocal asymmetric cloners. This method can be extended further in a similar fashion to produce more than three pairs. 

\section{ $1 \rightarrow 3$ broadcasting of entanglement using $1 \rightarrow 3$ asymmetric cloners}
\label{sec:1to3broadentdir}

In this section, we explore the idea of broadcasting to generate three entangled pairs, using optimal $1 \rightarrow 3$ asymmetric cloners (Eq.\eqref{eq:1to3cloner}). For this, we use non-maximally entangled state as our resource state (Eq.\eqref{eq:nme}).
\begin{figure}[t]
\centering
        \begin{subfigure}[b]{0.22\textwidth} \includegraphics[scale=0.25]{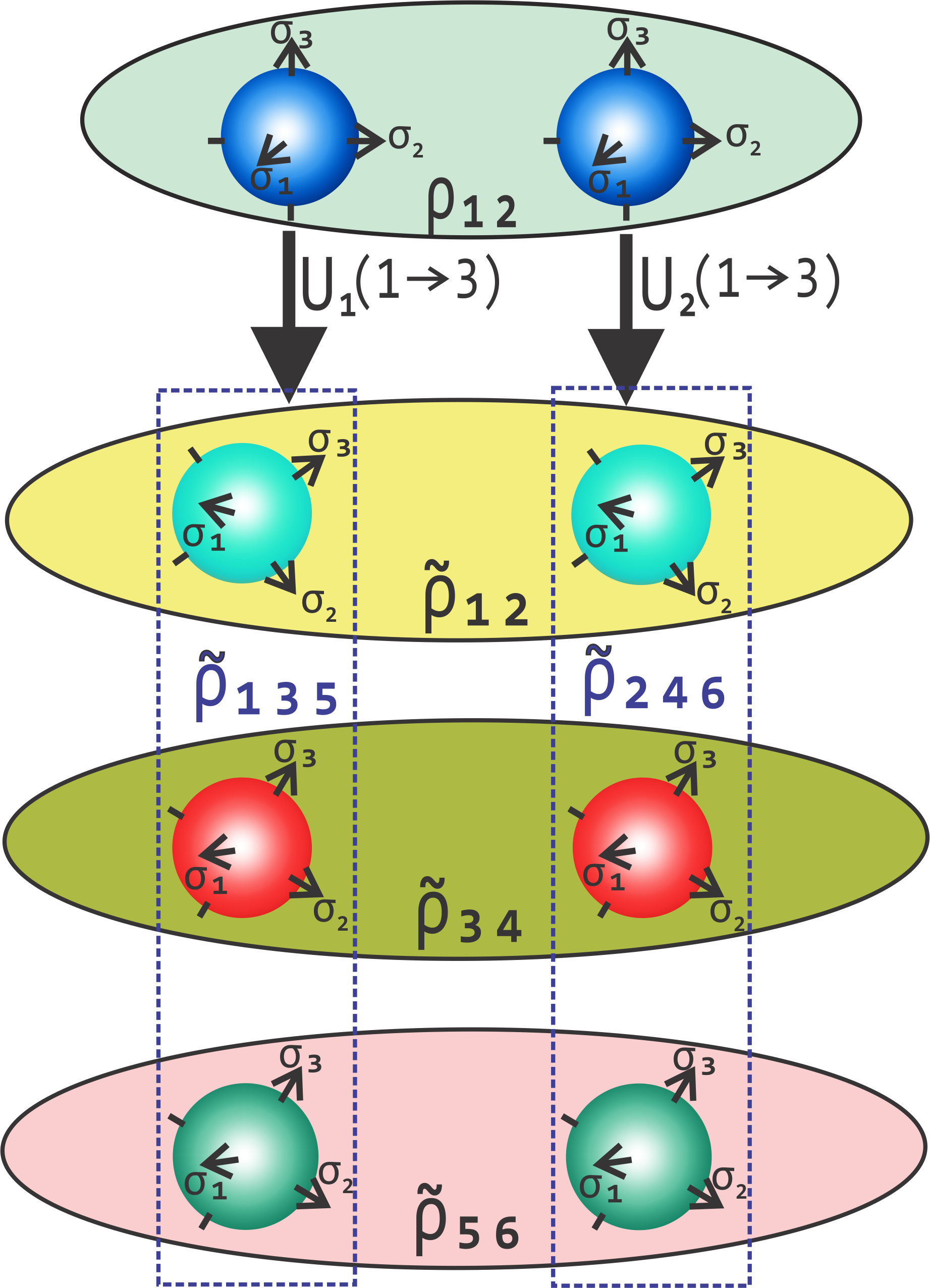}
                \caption{Local broadcasting}
                \label{fig:1to3directloc}
        \end{subfigure}\hspace{0.02\textwidth}%
        \begin{subfigure}[b]{0.23\textwidth}
            \includegraphics[scale=0.25]{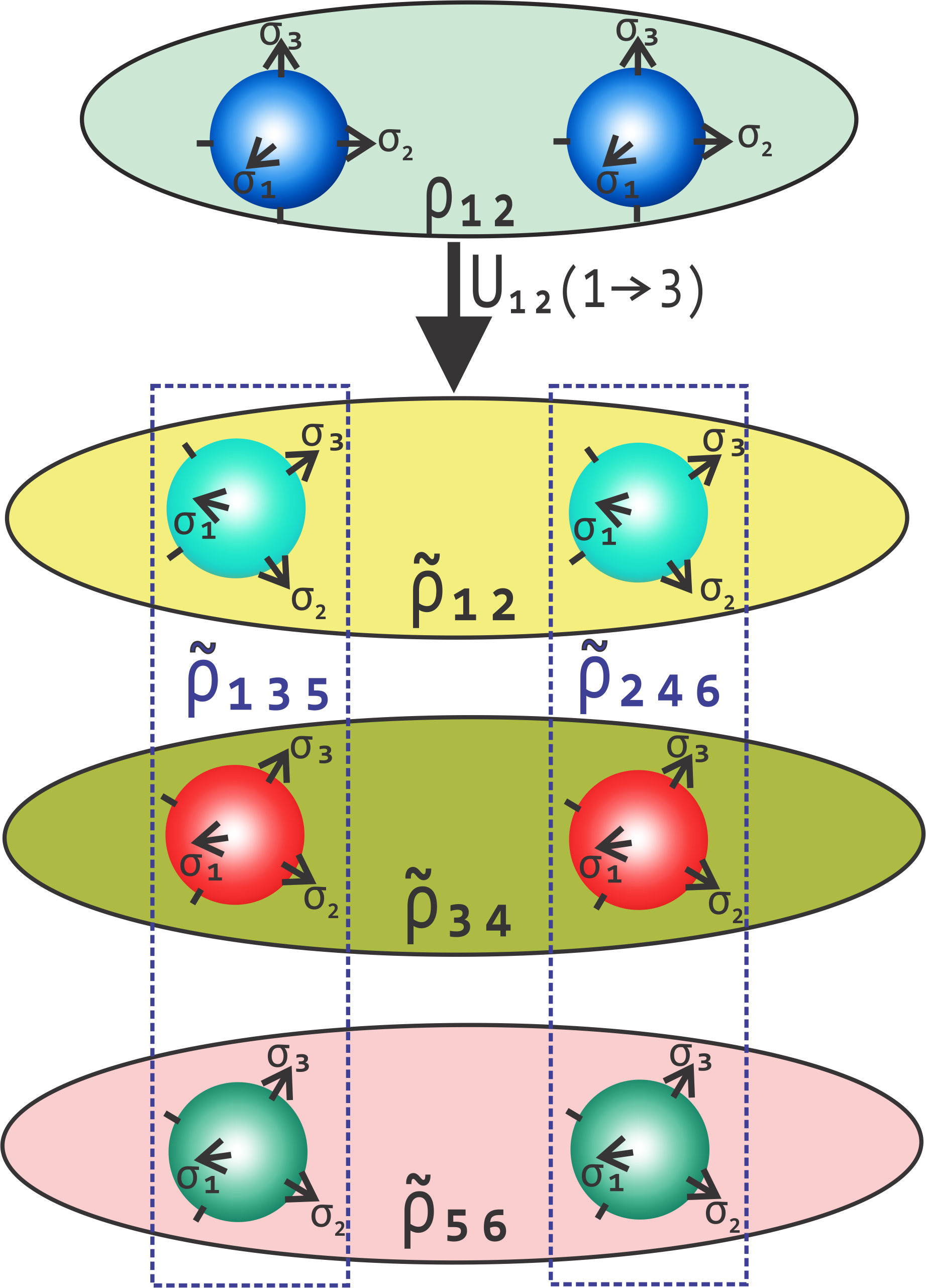}
                \caption{Nonlocal broadcasting}
                \label{fig:1to3directnloc}
        \end{subfigure}%
        \caption{The figures depict $1 \rightarrow 3$ asymmetric broadcasting of $\rho_{12}$ via direct $1$ to $3$ cloning. Only the horizontal nonlocal pairs ($\tilde{\rho}_{12}$, $\tilde{\rho}_{34}$ and $\tilde{\rho}_{56}$) have been highlighted for clarity.}\label{fig:1to3direct}
\end{figure}

\subsection{Broadcasting via local cloning}
We apply the cloner in Eq.\eqref{eq:1to3cloner} with $d=2$ on both the sides locally as shown in Fig (\ref{fig:1to3directloc}). After tracing out appropriate qubits, we obtain the nonlocal pairs represented by $\tilde\rho_{ij}$, where $i \in \{1,3,5\},j \in \{2,4,6\}$, and the local pairs obtained are given by $\tilde\rho_{kl}$, where $\{k,l\} \in \{\{1,3\},\{1,5\},\{3,5\},\{2,4\},\{4,6\},\{2,6\}\}$.

\noindent \begin{theorem}
\label{theo:broadloc1to3d}
\textit{Given a non-maximally entangled state $\ket{\psi_{12}}$, it is impossible to broadcast entanglement (optimally and non-optimally) into three lesser entangled states, using $1 \rightarrow 3$ optimal universal asymmetric cloning transformations locally. 
}
\end{theorem}
\noindent \emph{Proof:}
There are six possible groups of nonlocal pairs that are obtained at the end of the cloning transformations. One such group composed of horizontal nonlocal pairs is as follows :
\begin{equation}
\begin{split}
& \tilde{\rho}_{12} = \biggl\{ \{0,0,A_1B\}, \{0,0,A_1B\}, A_1^2\mathbb{T}^{N} \biggr\}, \\
& \tilde{\rho}_{34} = \biggl\{ \{0,0,A_2B\}, \{0,0,A_2B\}, A_2^2\mathbb{T}^{N} \biggr\}, \\
& \tilde{\rho}_{56} = \biggl\{ \{0,0,A_3B\}, \{0,0,A_3B\}, A_3^2\mathbb{T}^{N} \biggr\},\\
\end{split}
\end{equation}
where $A_1=(3 \alpha ^2+3 \alpha \beta +3 \alpha \gamma +\beta \gamma)/3$,$A_2=(3 \alpha \beta +\alpha \gamma +3 \beta ^2+3 \beta \gamma)/3$,$A_3=(\alpha \beta +3 \alpha \gamma +3 \beta \gamma +3 \gamma ^2)/3$,$B=(\alpha ^2+\alpha \beta +\alpha \gamma +\beta ^2+\beta \gamma +\gamma ^2)(2k-1)$, $\mathbb{T}^{N}$ is the correlation matrix of the initial input state (non-maximally entangled state) $\rho_{12}=\dyad{\psi_{12}}$. The expressions for the above states reduce to the nonlocal output state obtained in \cite{kar} for the symmetric case i.e. $\alpha=\beta=\gamma=\frac{1}{\sqrt{6}}$. 

By using \textbf{PH} criterion, we find that it is impossible to have all the pairs inseparable simultaneously. The procedure is repeated for all the other possible groups of nonlocal pairs and the inference is same in those cases. Hence, we arrive at a conclusion that given a non-maximally entangled state $\ket{\psi_{12}}$, it is impossible to broadcast entanglement (optimally and non-optimally) into three lesser entangled states, using $1 \rightarrow 3$ optimal universal asymmetric cloning transformations (Eq.\eqref{eq:1to3cloner}) locally.

\begin{figure}[b]
\centering
    \includegraphics[scale=0.5]{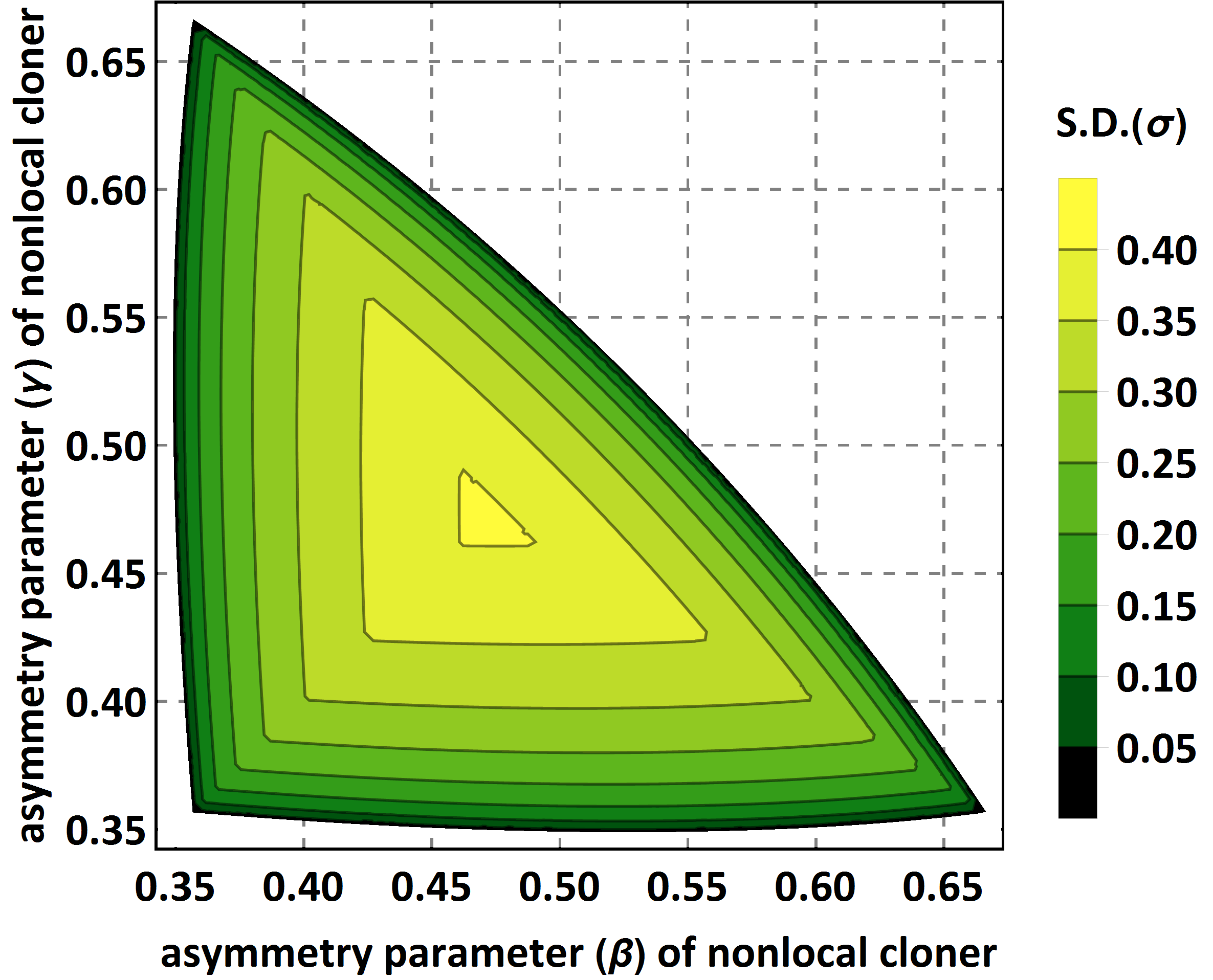} 
  \caption{\noindent \scriptsize The figure shows a contour plot of the allowed range of values of $k$ for a successful broadcasting operation, given the application of a $1 \rightarrow 3$ nonlocal cloner parameterised by $\beta$ and $\gamma$. The hue depicts the value of $\sigma$ such that the broadcasting range is $(0.5 - \sigma) \le k \le (0.5 + \sigma)$.}
\label{fig:succs1to3done}
\end{figure}

\subsection{Broadcasting via nonlocal cloning}
We start with the non-maximally entangled state again, and apply the cloner in Eq.\eqref{eq:1to3cloner} with $d=4$, as shown in Fig. (\ref{fig:1to3directnloc}). We obtain the nonlocal pairs represented by $\tilde\rho_{ij}$, where $i \in \{1,3,5\},j \in \{2,4,6\}$, and the local pairs obtained are then given by $\tilde\rho_{kl}$, where $\{k,l\} \in \{\{1,3\},\{1,5\},\{3,5\},\{2,4\},\{4,6\},\{2,6\}\}$.

Among all the possible groups of nonlocal output pairs, the following group composed of horizontal pairs were found to be inseparable :

\begin{align}
& \tilde{\rho}_{12} = \biggl\{ \{0,0,B_1\}, \{0,0,B_1\}, \frac{B_1}{(2k-1)}\mathbb{T}^{N} \biggr\}, \\
& \tilde{\rho}_{34} = \biggl\{ \{0,0,B_2\}, \{0,0,B_2\}, \frac{B_2}{(2k-1)}\mathbb{T}^{N} \biggr\}, \\
& \tilde{\rho}_{56} = \biggl\{ \{0,0,B_3\}, \{0,0,B_3\}, \frac{B_3}{(2k-1)}\mathbb{T}^{N} \biggr\}.
\end{align}

The above states reduce to the nonlocal output state obtained in \cite{kar} for the symmetric case i.e. $\alpha=\beta=\gamma=\frac{\sqrt{2}}{3}$.

The local output pairs obtained are,
\begin{align}
& \tilde{\rho}_{13} = \tilde{\rho}_{24} = \biggl\{ \{0,0,B_1\}, \{0,0,B_2\}, C_3\mathbb{I} \biggr\}, \\
& \tilde{\rho}_{35} = \tilde{\rho}_{46} = \biggl\{ \{0,0,B_2\}, \{0,0,B_3\}, C_1\mathbb{I} \biggr\}, \\
& \tilde{\rho}_{15} = \tilde{\rho}_{26} = \biggl\{ \{0,0,B_1\}, \{0,0,B_3\}, C_2\mathbb{I} \biggr\},
\end{align}
where $B_1=(10 \alpha ^2+5 \alpha \beta +5 \alpha \gamma +\beta \gamma)(2k-1)/10$, $B_2=(5 \alpha \beta +\alpha \gamma +10 \beta ^2+5 \beta \gamma)(2k-1)/10$, $B_3=(\alpha \beta +5 \alpha \gamma +5 \beta \gamma +10 \gamma ^2)(2k-1)/10$, $C_1=(2 \alpha ^2+\alpha \beta +\alpha \gamma +5 \beta \gamma)/10$, $C_2=(\alpha \beta +5 \alpha \gamma +2 \beta ^2+\beta \gamma)/10$, $C_3=(5 \alpha \beta +\alpha \gamma +\beta \gamma +2 \gamma ^2)/10$, and $\mathbb{T}^{N}$ is the correlation matrix of the initial input state $\rho_{NME}=\dyad{\psi_{12}}$.

We apply the \textbf{PH} criterion to determine the condition for which all these nonlocal output pairs will be inseparable, and all the local output pairs will be simultaneously separable. We obtain the broadcasting region involving three parameters : $k$ (input state parameter); $\beta,\gamma$ (asymmetric parameters of the cloner). In Fig. (\ref{fig:succs1to3done}) we plot the variation in broadcasting range with the asymmetry parameters of the cloner applied. The hue chart on the right of the figure depicts the values of $\sigma$, such that the range of allowed values of $k$ is $(0.5-\sigma) \le k \le (0.5+\sigma)$. We observe that for all the input states having $0.09 \le k \le 0.91$, there exists an asymmetric cloner parameterised by $\{\beta,\gamma\}$, which can be used to successfully broadcast the given state into three lesser entangled pairs optimally. The approximate ranges for the asymmetry parameters of the two cloner are $0.3 < \beta,\gamma < 0.7$. The maximum range is achieved for a perfectly symmetric cloner ($\alpha=\beta=\gamma=\frac{\sqrt{2}}{3}\approx0.47$). The range becomes narrower as the asymmetry of the cloner increases i.e. we move away from the point of symmetry in the plot. 

On comparing the two strategies for nonlocal broadcasting, to generate three entangled pairs out of one entangled pair, we find that using $1 \rightarrow 3$ cloner does a better job than successive strategy. It is able to broadcast a wider range of input state space. The number of ancillary qubits needed in both the strategies remain same. One can choose the strategy depending on the experimental challenges in construction of the two different kinds of cloning machines ($1 \rightarrow 2$ and $1 \rightarrow 3$). The implementation of $1 \rightarrow 2$ cloners is expected to be simpler than $1 \rightarrow 3$ cloners and there the approach of successive broadcasting could be beneficial over using direct $1 \rightarrow 3$ cloners. The successive approach can be extended to produce more than three copies, and helps to avoid the need to build $1 \rightarrow M$ cloners in general for a large part of the input state space. 

The limit on the number of entangled pairs that can be obtained from nonlocal broadcasting using symmetric cloners is six copies \cite{kar}. Symmetric cloning distributes the entanglement equally among all the nonlocal pairs. However, it is possible to distribute different amounts of entanglement to different nonlocal pairs via asymmetric cloning. Using the successive strategy, one can continue to apply an asymmetric cloner on the pair that has the highest amount of entanglement at each step, to create more entangled pairs. The flexibility of creating pairs with different amounts of entanglement might help in surpassing the limit on the number of copies set by symmetric cloning. It is also possible to use a direct $1 \rightarrow M$ asymmetric cloner and obtain nonlocal pairs with varying amounts of entanglement. Therefore, we strongly conjecture that the limit of six copies for broadcasting of entanglement via symmetric nonlocal cloning can be surpassed using any of the two strategies discussed above, both of which involve asymmetric cloning.

\section{ Broadcasting using arbitrary unitaries}
\label{sec:arbunit}
In this section, we deal with the problem of local broadcasting of entanglement independent of the use of standard cloning machines. We show that there exist arbitrary unitary operations using which we can achieve broadcasting for a wider range of input states, than what can be best obtained by using the cloning operation. We demonstrate this using two different classes of resource states, namely Werner-like and Bell diagonal states. Unlike cloning, we present an economical way of accomplishing the task in the sense that it does not employ any ancillary qubits. 

We start with two parties A and B sharing an entangled input state $\rho_{12}$, with qubits numbered $1$ and $2$ on A's side and B's side respectively. An arbitrary local unitary operation $U_b$ (say) is applied on pairs of qubits numbered $(1,3)$ and $(2,4)$ separately. The diagonal pair of nonlocal output states obtained are, 
\begin{equation}
\begin{split}
& \tilde{\rho}_{14}=Tr_{23}[U_b\otimes U_b (\rho_{12}\otimes \sigma_{34})U_b^{\dagger}\otimes U_b^{\dagger}]\\
& \tilde{\rho}_{23} =Tr_{14}[U_b\otimes U_b (\rho_{12}\otimes \sigma_{34})U_b^{\dagger}\otimes U_b^{\dagger}]
\end{split}    
\end{equation}
where $\sigma_{34} = \dyad{00}$ and $U_b$ is an arbitrary unitary matrix of dimension $4$. 

Now we apply the \textbf{PH} criterion by calculating the appropriate determinants (as in Eq.~\eqref{eq:w3w4}) to check if the diagonal pairs obtained above are simultaneously entangled. To find the unitary that maximises the broadcasting range, we use the parameterised version of the unitary \cite{Hedemann}, $U_{4}$ given by,
\begin{equation}%
\begin{array}{l}
 \left(\!\! {\begin{array}{*{20}c}
   a & {bc} & {bde} & {bdf}  \\
   {b^* g} & {\left(\!\! \begin{array}{l}
  - a^* cg \\ 
  + d^* hm \\ 
 \end{array}\!\! \right)} & {\left(\!\! \begin{array}{l}
  - a^* deg  \\ 
  - c^* ehm \\ 
  + f^* hn \\ 
 \end{array}\!\! \right)} & {\left(\!\! \begin{array}{l}
  - a^* dfg \\ 
  - c^* fhm \\ 
  - e^* hn \\ 
 \end{array}\!\! \right)}  \\
   {b^* h^* j} & {\left(\!\! \begin{array}{l}
  - a^* ch^* j \\ 
  - d^* g^* jm \\ 
  + d^* l^* n^*  \\ 
 \end{array}\!\! \right)} & {\left(\!\! \begin{array}{l}
  - a^* deh^* j \\ 
  + c^* eg^* jm \\ 
  - c^* el^* n^*  \\ 
  - f^* g^* jn \\ 
  - f^* l^* m^*  \\ 
 \end{array}\!\!\! \right)} & {\left(\!\! \begin{array}{l}
  - a^* dfh^* j \\ 
  + c^* fg^* jm \\ 
  - c^* fl^* n^*  \\ 
  + e^* g^* jn \\ 
  + e^* l^* m^*  \\ 
 \end{array}\!\!\! \right)}  \\
   {b^* h^* l} & {\left(\!\! \begin{array}{l}
  - a^* ch^* l \\ 
  - d^* g^* lm \\ 
  - d^* j^* n^*  \\ 
 \end{array}\!\! \right)} & {\left(\!\! \begin{array}{l}
  - a^* deh^* l \\ 
  + c^* eg^* lm \\ 
  + c^* ej^* n^*  \\ 
  - f^* g^* ln \\ 
  + f^* j^* m^*  \\ 
 \end{array}\!\!\! \right)} & {\left(\!\! \begin{array}{l}
  - a^* dfh^* l \\ 
  + c^* fg^* lm \\ 
  + c^* fj^* n^*  \\ 
  + e^* g^* ln \\ 
  - e^* j^* m^*  \\ 
 \end{array}\!\!\! \right)}  \\
\end{array}}\!\!\! \right) \\ 
 \end{array}\!\!,
\label{eq:parametrized_unitary}
\end{equation}
where $|a|^2+|b|^2=1,|c|^2+|d|^2=1,|e|^2+|f|^2=1,|g|^2+|h|^2=1,|j|^2+|l|^2=1,|m|^2+|n|^2=1$.
We substitute the entries of the matrix $U_b$ with the ones in $U_4$ such that $U_b=U_4$. We now perform a randomised numerical search over the parameters \{a,b,c,d,e,f,g,h,j,l,m,n\} and maximise the range of broadcasting. The parameters are assumed to be real for simplifying the search. We demonstrate the method using two classes of mixed entangled resource states namely Werner-like states and Bell-diagonal states. 

\begin{center}\noindent\textit{Werner-like states}\end{center}
Werner-like states \cite{wer1} are represented canonically as,
\begin{equation}
\rho^w_{12} = \{\vec{x}^w,\vec{x}^w,\mathbb{T}^{w}\},
\end{equation}
where $\vec{x}^w=\{0,0,p(2k-1)\}$ is the Bloch vector and $\mathbb{T}^{w}=\textit{diag}\{2p\sqrt{k(1-k)},-2p\sqrt{k(1-k)},p\}$ is the correlation matrix. Here, $k$ and $p$ are state parameters satisfying $0 \le k,p \le 1$. As a side note, this class of state reduces to the non-maximally entangled state for $p=1$. 

We apply the procedure described above with $\rho_{12}=\rho^w_{12}$ as the resource state and find a unitary that maximises the broadcasting range. An approximate representation of the unitary (rounded to 4 decimal places) is as follows,

\begin{equation}
U^{w}_{opt} = 
\begin{pmatrix}
 1 & 0 & 0 & 0 \\
 0 & -0.0773 & -0.9898 & -0.1194 \\
 0 & -0.8255 & 0.1306 & -0.5490 \\
 0 & 0.5590 & 0.0561 & -0.8272 \\
\end{pmatrix}
\label{eq:optunitarywer}
\end{equation}
which is obtained at the parameter configuration given by, 
\begin{equation}
\begin{split}
& \{a,b,c,d,e,f,g,h,j,l,m,n\}\\
= &\cos{\{0,\frac{\pi}{2},\frac{\pi}{5},\frac{3\pi}{10},\frac{2\pi}{5},\frac{\pi}{10},\frac{2\pi}{5},\frac{\pi}{10},0,\frac{\pi}{2},\frac{2\pi}{5},\frac{9\pi}{10}\}}.
\end{split}
\label{eq:wer_par}
\end{equation}
where $\cos{\{x_1,x_2,\dots\}}=\{\cos{(x_1)},\cos{(x_2)},\dots\}$. 

A more accurate representation of the unitary can be obtained by substituting the values of the parameters from Eq\eqref{eq:wer_par} in Eq\eqref{eq:parametrized_unitary}. 

\begin{figure}[h]
\centering
    \includegraphics[scale=1]{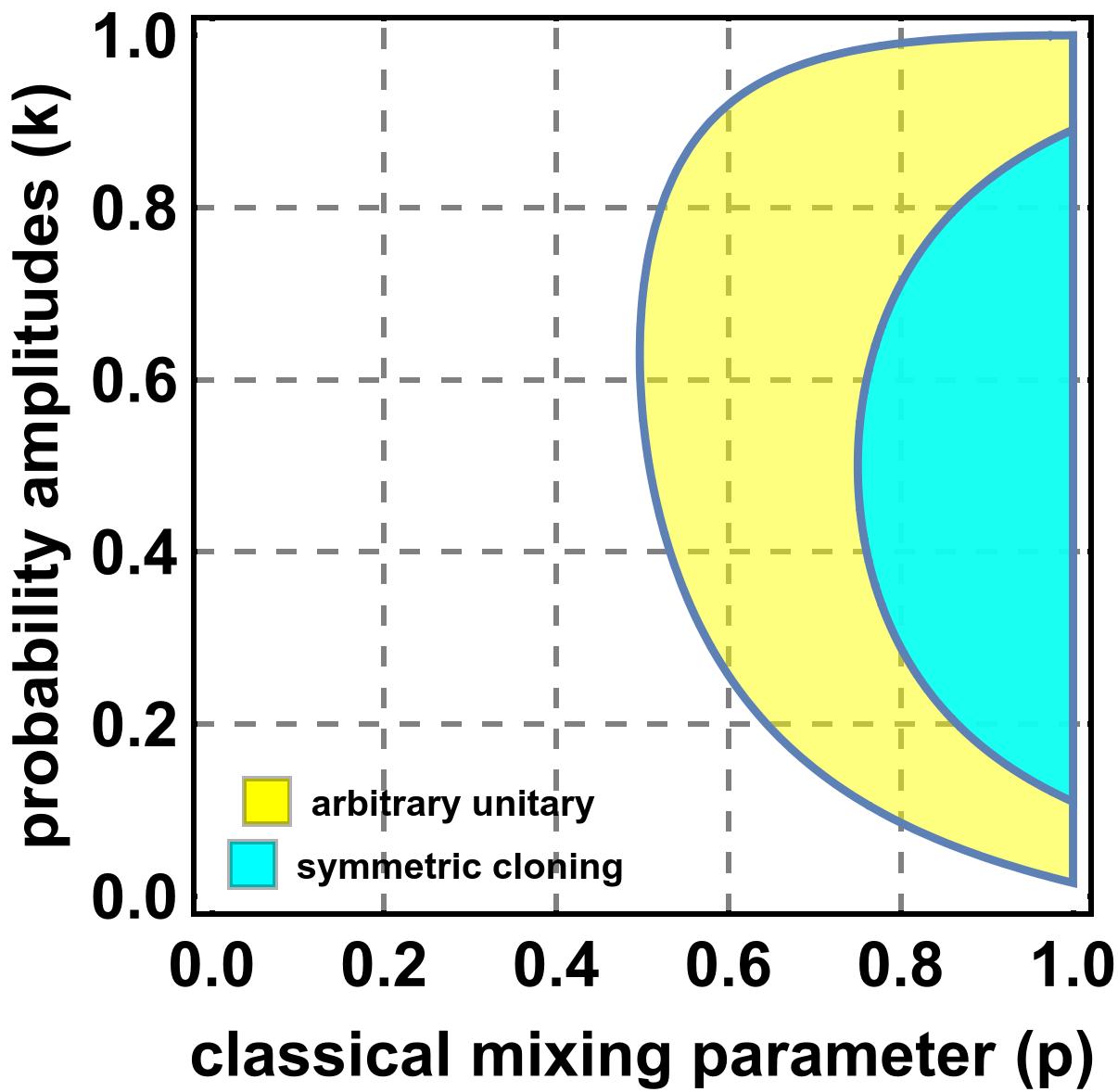}
  \caption{\scriptsize The figure shows the broadcastable region for Werner-like states when using arbitrary unitary and symmetric cloner. The yellow (light gray) region of input state space covered by the arbitrary unitary is strictly larger than and engulfs the cyan (dark gray) region covered by the symmetric Buzek-Hillary (B-H) cloner.}
\label{fig:wer_arbit}
\end{figure}

In Fig~\ref{fig:wer_arbit}, we plot the region of input state space for which broadcasting can be achieved using the arbitrary unitary and contrast it with the performance of the best possible cloner. The unitary in Eq.\eqref{eq:optunitarywer} beats the performance obtained using Buzek-Hillary (B-H) cloner. It is able to broadcast a larger set of input states which includes the set of states broadcastable using the cloner. This enables us to achieve broadcasting from weaker entangled states, something which was other impossible to achieve using cloners.

\begin{center}\noindent\textit{Bell-diagonal states}\end{center}
Bell diagonal states \cite{bd1} are a larger class of states represented canonically as,
\begin{equation}
\rho^{bd}_{12} = \{\vec{0},\vec{0},\mathbb{T}^{bd}\},
\end{equation}
where $\vec{0}$ is the Bloch vector and $\mathbb{T}^{bd}=\textit{diag}\{c_1,c_2,c_3\}$ is the correlation matrix, with $-1 \le c_i \le 1$. This state can also be written as a combination of the four Bell states as $\rho^{bd}_{12} = \sum_{u,v} \lambda_{uv} \dyad{\gamma_{uv}}$, where the four Bell states $\ket{\gamma_{uv}}\equiv (\ket{0,v}+(-1)^u \ket{1,1 \oplus v})/\sqrt{2}$ represent the eigenstates of $\rho^{bd}_{12}$ and $u,v\in \{0,1\}$. The eigenvalues are given by,
\begin{equation*}
    \lambda_{uv} = \frac{1}{4} [1 + (-1)^u c_1 - (-1)^{(u+v)}c_2 + (-1)^v c_3].
\end{equation*}
It is required that $\lambda_{uv} \ge 0$ for $\rho^{bd}_{12}$ to be a valid density operator. 

We repeat the procedure described above with $\rho_{12}=\rho^{bd}_{12}$ as our resource state and find a unitary that maximises the broadcasting range. An approximate representation of the unitary (rounded to 4 decimal places) is as follows,

\begin{equation}
U^{bd}_{opt} = 
\begin{pmatrix}
  0.8090 & 0.1816 & -0.4523 & 0.3286 \\
 -0.1816 & -0.8273 & -0.4301 & 0.3125 \\
 0.5590 & -0.5317 & 0.5148 & -0.3740 \\
 0 & 0 & -0.5878 & -0.8090 \\
\end{pmatrix}
\label{eq:optunitarybd}
\end{equation}
which is obtained at the parameter configuration given by, 
\begin{equation}
\begin{split}
& \{a,b,c,d,e,f,g,h,j,l,m,n\}\\
= &\cos{\{\frac{\pi}{5},\frac{7\pi}{10},\frac{3\pi}{5},\frac{9\pi}{10},\frac{4\pi}{5},\frac{3\pi}{10},\frac{2\pi}{5},\frac{9\pi}{10},0,\frac{\pi}{2},\pi,\frac{\pi}{2}\}}.
\end{split}
\label{eq:bd_par}
\end{equation}
where $\cos{\{x_1,x_2,\dots\}}=\{\cos{(x_1)},\cos{(x_2)},\dots\}$. 

A more accurate representation of the unitary can be obtained by substituting the values of the parameters from Eq\eqref{eq:bd_par} in Eq\eqref{eq:parametrized_unitary}. 

\begin{figure}[h]
\centering
    \includegraphics[scale=0.65]{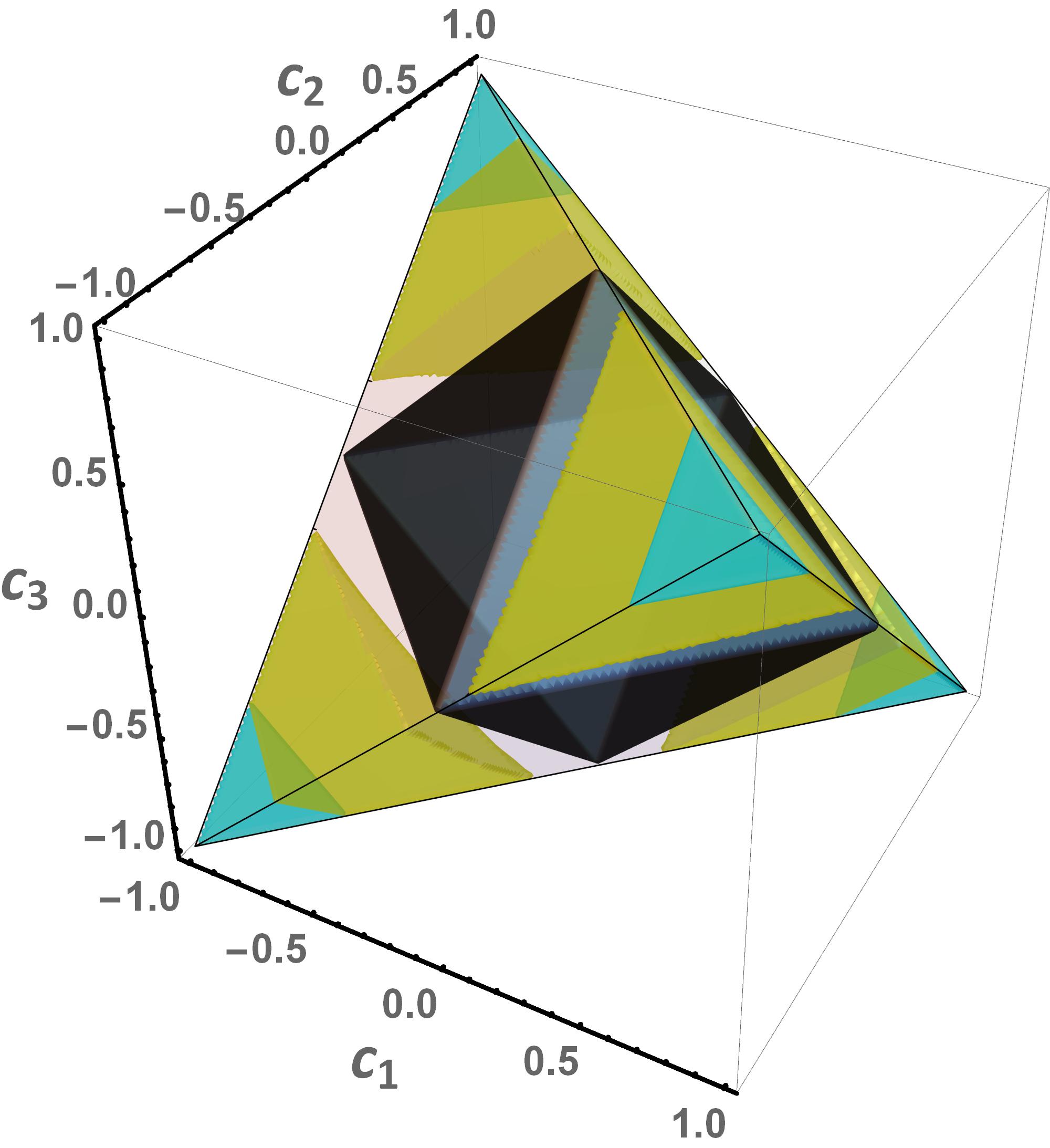}
  \caption{\scriptsize The figure shows the state space for Bell-diagonal states. At the vertex tuples (-1,-1,-1), (1,1,-1), (1,-1,1) and (-1,1,1) of the translucent tetrahedron lie the Bell states $\ket{\gamma_{uv}}$. The cones emerging from these corners depict the broadcastable regions. The small (blue) cones which are associated with the use of symmetric Buzek-Hillary (B-H) cloner are engulfed by the large (chrome-yellow) cones which correspond to the use of arbitrary unitary. The (black) octahedron in the middle of the tetrahedron depicts the separable region within the Bell-diagonal state space.}
\label{fig:bd_arbit}
\end{figure}

In Fig~\ref{fig:bd_arbit}, we plot the region of input state space for which broadcasting can be achieved using the arbitrary unitary and contrast it with the performance of the best possible cloner. The smaller and larger cones at the corners of the tetrahedron depict the broadcastable regions when using the Buzek-Hillary (B-H) cloner and the arbitrary unitary respectively. As can be observed from the figure, using the arbitrary unitary (Eq.\eqref{eq:optunitarybd}) enhances the range of states which can be broadcast significantly.  

A notable feature of this economical method is that it is difficult to achieve optimal broadcasting as defined in the literature i.e. the local output pairs should not be entangled at the end of broadcasting. The total initial entanglement in this protocol gets redistributed in the local output pairs (same side) and non-local output pairs (opposite sides). Since there are no ancilla qubits, it is difficult to redistribute in a way such that the local pairs are completely separable.
In cloning based methods, some of the entanglement is shared with the ancilla qubits which is also inaccessible in terms of usability, but helps in achieving optimal broadcasting as defined in the literature. We argue that both these scenarios are equivalent i.e. they both distribute entanglement in pairs other than nonlocal output pairs that cannot be utilised directly.

By this we present a way to achieve broadcasting by applying special local unitaries. One can implement these unitaries using efficient methods available in literature \cite{bullock,zhangcir,shende}. Some of these methods use as small as $23$ elementary gates, of which at most $4$ (CNOT) entail multi-qubit interactions \cite{bullock}. This clearly shows the existence of physically realizable unitaries which do a significantly better job than the traditional pathway of achieving broadcasting via local cloning. This also opens up a new domain of investigation for broadcasting of entanglement (correlation in general). One can also think of using better optimization methods such as gradient-ascent as in \cite{grape} or Nelder-Mead method \cite{nm} rather than a coarse search over the parameter space for finding such unitaries, if the range maximisation can be transformed into optimizing a single real valued function.

\section{Conclusion}
\label{sec:conclusion}

In a nutshell, the work gives an exhaustive description and analysis of broadcasting of entanglement and correlations beyond entanglement (discord) using asymmetric cloners. In this work we have particularly proposed a new method called successive broadcasting which helps in generating more than two copies of entangled pairs. We also introduce a novel direction of broadcasting using arbitrary unitaries. Such a direction differs from the traditional methods of cloning. We show via numerical examples that there exist special unitaries that outperform all the cloning machines in broadcasting entanglement. This brings out the fundamental fact that broadcasting of entanglement is not limited by cloning. Although cloning helps to achieve broadcasting, it might not be the optimal approach.

Specifically, we investigate $1 \rightarrow 2$ broadcasting of entanglement and correlations that go beyond entanglement (discord) using optimal asymmetric Pauli cloners and most general two-qubit mixed state as a resource state. We exemplify our result with the help of Maximally Entangled Mixed States (MEMS) and show the variation of broadcasting range with the asymmetry parameter of the cloning machine. This example also gives us a better understanding on how the broadcasting range depends upon the initial amount of entanglement. We also give proofs that it is impossible to broadcast quantum correlations beyond entanglement optimally with these cloners, which was hypothesized in \cite{sourav}. In this work, we also address the problem of $1 \rightarrow 3$ broadcasting of entanglement with non-maximally entangled (NME) state as a resource. We adopt two strategies of cloning for this purpose. For nonlocal broadcasting, we show that $1 \rightarrow 3$ optimal asymmetric cloners always perform better than the successive broadcasting technique. However, the successive strategy carries importance in the absence of $1 \rightarrow 3$ cloners. We find that better broadcasting range can be achieved when an asymmetric cloner is paired with a symmetric cloner for $1 \rightarrow 3$ broadcasting of entanglement using successive strategy, as compared to successive use of two symmetric cloners. We also prove that it is impossible to broadcast entanglement locally using both the strategies. The possibility of creating more than six copies of entangled pairs using nonlocal asymmetric cloning remains an interesting open question. Our hypothesis is that it is feasible using any of the two strategies discussed herein. It would also be interesting to explore all the above cases in more generality i.e. where different asymmetric cloners are applied by the two parties that share the input state. Finally, we introduce the notion of broadcasting independent of cloning. This is a first step in the direction towards characterizing the best possible unitaries for the task of broadcasting of entanglement. We show the existence of local unitary operations which can be applied to get a better broadcasting range than what is best obtained by standard cloning operations, while being economical as well. This can be generalised for other broader classes of input states by employing clever search techniques and would be a good topic for future exploration.
\subsection*{Acknowledgement} A. Jain thanks Mr. Erik Anderson for useful discussions on the hyperspherical parameterization of unitary matrices. The authors thank Mr. Rounak Mundra for insightful discussion on calculation of geometric discord.
\bibliography{apssamp}

\begin{thebibliography}{65}
\bibitem [{foo()}]{foot-color}%
  \BibitemOpen
  \href@noop {} {\bibinfo  {journal} {The difference in color and orientation
  of the qubits including the states have been introduced to denote the
  non-identical outputs from the asymmetric cloner. However, when the input is
  a non-maximally entangled state, the outputs on A's side are same as that on
  B's side.}\ }\BibitemShut {NoStop}%
\end{thebibliography}
\end{document}